\documentclass[fleqn,usenatbib,useAMS]{mnras}

\usepackage{graphicx}	% Including figure files
\usepackage{amsmath}	% Advanced maths commands
\usepackage{amssymb}	% Extra maths symbols
\usepackage{multicol}        % Multi-column entries in tables
\usepackage{bm}		% Bold maths symbols, including upright Greek
\usepackage{pdflscape}	% Landscape pages
\usepackage{rotating}
\usepackage{geometry}
\usepackage{longtable}
\def\lea{\mathrel{<\kern-1.0em\lower0.9ex\hbox{$\sim$}}}
\def\gea{\mathrel{>\kern-1.0em\lower0.9ex\hbox{$\sim$}}}

\usepackage[T1]{fontenc}
\usepackage{ae,aecompl}

\usepackage{newtxtext,newtxmath}
\title[Improving Cluster Age Estimates]{
Improving Star Cluster Age Estimates in PHANGS-HST Galaxies and 
the Impact on Cluster Demographics in NGC 628
}

\author[Whitmore et al.]
{Bradley~C.~Whitmore$^{1}$,\thanks{Contact e-mail:\href{mailto:whitmore@stsci.edu}{whitmore@stsci.edu}}
Rupali~Chandar,$^{2}$
Janice~C.~Lee,$^{3}$
Matthew~Floyd,$^{2}$ 
Sinan~Deger,$^{4}$
\newauthor
James Lilly,$^{5}$
Rebecca~Minsley,$^{6}$
David~A.~Thilker,$^{7}$
M\'ed\'eric~Boquien,$^{8}$
\newauthor
Daniel~A.~Dale,$^{5}$
Kiana Henny,$^{5}$
Fabian Scheuermann,$^{9}$
Ashley~T.~Barnes,$^{10}$
\newauthor
Frank Bigiel,$^{10}$
Eric~Emsellem,$^{11}$
Simon~Glover,$^{12}$
Kathryn~Grasha,$^{13}$
Brent~Groves,$^{14,15}$
\newauthor
Stephen~Hannon,$^{15}$
Ralf~S.~Klessen,$^{12,16}$
Kathryn~Kreckel,$^{9}$ 
J.~M.~Diederik~Kruijssen,$^{17}$
\newauthor
Kirsten~L.~Larson,$^{1}$
Adam~Leroy,$^{18}$
Angus~Mok,$^{19}$
Hsi-An Pan,$^{20}$
Francesca~Pinna,$^{21}$
\newauthor
Patricia~Sánchez-Blázquez,$^{22,23}$
Eva~Schinnerer,$^{21}$
Mattia~C.~Sormani,$^{12}$
Elizabeth~Watkins,$^{9}$
\newauthor
Thomas Williams$^{9}$
\\
\bigskip \\
$^{1}$Space Telescope Science Institute, 3700 San Martin Drive, Baltimore, MD, USA\\
$^{2}$Ritter Astrophysical Research Center, University of Toledo, Toledo, OH 43606, USA\\
$^{3}$Gemini Observatory/NSF’s NOIRLab, 950 N. Cherry Avenue, Tucson, AZ, USA\\
$^{4}$TAPIR, California Institute of Technology, Pasadena, CA, 91125 USA\\
$^{5}$Department of Physics \& Astronomy, University of Wyoming, Laramie, WY 82071, USA \\
$^{6}$Department of Physics and Astronomy, University of Arizona, USA\\
$^{7}$Department of Physics and Astronomy, The Johns Hopkins University, Baltimore, MD, 21218 USA\\
$^{8}$Centro de Astronom\'ia, (CITEVA), Universidad de Antofagasta,  Avenida Angamos 601, Antofagasta 1270300, Chile\\
$^{9}$Astronomisches Rechen-Institut, Zentrum f\" ur Astronomie der Universit\"at Heidelberg, M\"onchhofstra\ss e 12-14, D-69120 Heidelberg, Germany\\
$^{10}$Argelander-Institut f\"ur Astronomie, Universit\"at Bonn, Auf dem H\"ugel 71, 53121 Bonn, Germany\\
$^{11}$ European Southern Observatory,  Karl-Schwarzschild Str. 2, 85748 Garching bei Muenchen, Germany\\
$^{12}$Universit\"at Heidelberg, Zentrum f\"ur Astronomie, Institut f\"ur Theoretische Astrophysik, Heidelberg, Germany\\
$^{13}$Research School of Astronomy and Astrophysics, Australian National University, Canberra, ACT 2611, Australia\\
$^{14}$International Centre for Radio Astronomy Research, University of Western Australia, 35 Stirling Hwy, Crawley, WA 6009, Australia \\
$^{15}$Department of Physics and Astronomy, University of California, Riverside, CA, 92521 USA\\
$^{16}$Universit\"at Heidelberg, Interdisziplin\"ares Zentrum f\"ur Wissenschaftliches Rechnen, Heidelberg, Germany\\
$^{17}$Cosmic Origins Of Life (COOL) Research DAO, coolresearch.io\\
$^{18}$Department of Astronomy, The Ohio State University, 140 West 18th Ave., Columbus, OH 43210, USA\\
$^{19}$OCAD University, Toronto, Ontario, M5T 1W1, Canada\\
$^{20}$Department of Physics, Tamkang University, No.151, Yingzhuan Road, Tamsui District, New Taipei City 251301, Taiwan \\
$^{21}$Max Planck Institut f\"ur Astronomie, K\"onigstuhl 17, 69117 Heidelberg, Germany\\
$^{22}$Departamento de F\'isica de la Tierra y Astrof\'isica, Universidad Complutense de Madrid, E-28040 Madrid, Spain\\
$^{23}$Instituto de F\'isica de Partículas y del Cosmos IPARCOS, Facultad de Ciencias F\'isicas, Universidad Complutense de Madrid, 28040, Madrid, Spain\\
}

\date{These dates will be filled out by the publisher}

\pubyear{2021}

\begin{document}
\label{firstpage}
\pagerange{\pageref{firstpage}--\pageref{lastpage}}
\maketitle

% NOTE WHICH APPLIES THROUGH

\clearpage

\begin{abstract}

A long-standing problem when deriving the physical properties of stellar populations 
is the degeneracy between age, reddening, and metallicity. 
When a single metallicity is used
for all star clusters in a galaxy, this degeneracy can result in "catastrophic" errors for old globular clusters. 
Typically, approximately 10 -- 20 \% of all clusters detected in spiral galaxies can have ages that are incorrect by a factor of ten or more. 
In this paper we present 
a pilot study for four galaxies (NGC~628, NGC~1433, NGC~1365, and NGC~3351) from the PHANGS-HST survey. We describe 
methods to correct the age-dating for old globular clusters, by first identifying candidates 
using their colors, and then reassigning ages and reddening  based on a lower metallicity  solution. 
We find that young `Interlopers' can be identified from their  H$\alpha$ flux. CO (2-1) intensity or the presence of dust can also be used, but our tests show that they do not work as well. Improvements in the success fraction are possible at the $\approx15$ \% level (reducing the fraction of catastrophic age-estimates from  between 13 - 21 \% to 3 - 8 \%). 
A large fraction of the incorrectly age-dated globular clusters are systematically given ages around 100 Myr, polluting the  younger populations as well. 
Incorrectly age-dated globular clusters significantly impact the observed cluster age distribution in NGC~628, which affects the physical interpretation of cluster disruption in this galaxy.
For NGC~1365, we also demonstrate how to fix a second major age-dating problem, 
where very dusty young clusters with $E(B-V)>1.5$~mag are assigned old, globular-cluster like ages. 
Finally, we note the discovery of a dense population of $\approx300$ Myr clusters around the central region of NGC~1365.
and discuss how this results naturally from the dynamics in a barred galaxy.

\end{abstract}


%%%%%%%%%%%%%%%%% BODY OF PAPER %%%%%%%%%%%%%%%%%%

\section{Introduction}

% why its important to be able to age-date clusters
Estimating the ages of star clusters provides direct insight into the star formation process and the structure and evolution of galaxies. Clusters provide clocks for measuring a number of physical mechanisms that are fundamental to understand how galaxies operate, such as the timescales for giant molecular clouds to form star clusters, the timescales for feedback to stop star formation and hence conserve gas for the longer term evolution of the galaxy, and the timescales for the disruption of star clusters as  they return their member stars into the field population of the galaxy (\citealp{lada03},  \citealp{chandar06}, \citealp{whitmore07}, \citealp{kawamura09}, 
\citealp{bastian12a}, \citealp{fall12}, \citealp{whitmore14}, \citealp{grasha15}, \citealp{kruijssen19}, \citealp{chevance20}, \citealp{barnes20}, \citealp{kim22}).
%\cite{zabel22}).

A variety of increasingly sophisticated stellar population models have been developed to predict the observed color evolution and ages of clusters. These include models by
\cite{charlot91}, \cite{leitherer99}, \cite{bruzual03},  \cite{Vazdekis10}, \cite{maraston11},\ \cite{zackrisson11}, \cite{krumholz15}, and \cite{eldridge17}.

Similarly, a variety of increasingly sophisticated %software programs 
methods have been developed to age-date clusters, including: isochrone synthesis - \cite{charlot91}, reddening-free Q parameters - \citet{whitmore99}, hybrid color-color diagrams - \citet{whitmore02},  maximum likelihood fits -  \citet{chandar10a}, \citet{adamo17}, stochastically sampled probabalistic models -  
\cite{fouesneau10}, \cite{krumholz15}, \citet{ashworth17} and Bayesian analysis -  \citet{Boquien19}. 
Each of these approaches has its own strengths and weaknesses.
%have their own pros and cons. 
%%% 
In general, the results from different approaches agree reasonably well
%agreement between different approaches are reasonably good 
(e.g., \citealp{degrijs05}).
On the near horizon is an
increase in the wavelength range and sensitivity 
%by using
promised by JWST observations, which will allow us to identify and study the embedded cluster population and improve our ability to accurately age-date clusters.

However, all of these models and approaches share a common problem, which in many cases
introduces major systematic uncertainties which can seriously affect the science results. This is the long-standing age/metallicity/reddening degeneracy problem. The basic reason for the degeneracy is that integrated stellar populations can change from blue to red colors for three independent reasons; the age increases, reddening due to dust increases, or the metallicity increases. In a color-color diagram, this problem is seen by the fact that the predicted colors of aging clusters move from the blue (young) corner to the red (old) corner along a roughly diagonal line that  has about the same slope as the direction that reddening moves clusters in this space (e.g., \citealp{whitmore02}, \citealp{chandar04}, \citealp{smith07}, \citealp{Cockcroft11}, \citealp{forbes22}). 

To help break the degeneracies, people use several (generally three or more) photometric bands, which add structure to the shape of the Spectral Energy Distribution (SED) and help distinguish it from a straight line. The addition of the U-band is particularly important for this  (\citealt{anders04a}, \citealt{smith06}, \citealt{smith07}). 
In many cases, follow-up spectroscopy provides more robust age estimates than is possible from photometry alone (e.g., \citealt{chandar04}, \citealt{annibali18}, \citealt{whitmore20}, \citealt{forbes22}).
Other observational measurements (e.g., H$\alpha$ flux) can also help break the degeneracies  (e.g., \citealp{whitmore02},
\citealp{anders03},   \citealp{chandar10a}, \citealp{fouesneau12}, \citealp{ashworth17}, \citealp{barnes21}).

% However, 
In many cases the techniques used to help break the degeneracies are not enough,  as demonstrated  in \cite{whitmore20} for the LEGUS (Legacy Extragalactic UV Survey -  \citealp{calzetti15a}) galaxy NGC4449. 
% The most important systematic age-dating problem for most studies of cluster populations in star-forming galaxies occurs for the population of old globular clusters, 
By comparing ages estimated from spectra of verified old globular clusters, with those estimated using a standard SED-fitting approach, the authors found that ages for most old globular clusters were underestimated by a factor of 50 to 1000.  Instead of giving age estimates around 10 Gyr ages, most of the globular clusters were assigned ages between 5  and 500 Myr. In the current paper, we show that this problem can be  understood by looking at the shape of Simple Stellar Population (SSP) models in a color-color diagram. 
Other recent papers that  discuss this problem for PHANGS-HST (Physics at High Angular Resolution in Nearby Galaxies with the Hubble Space Telescope project; PI: J.C. Lee, GO-15654) and LEGUS galaxies are % \cite{whitmore20},
\cite{turner21}, 
\cite{deger22},
and \cite{hannon22}. This problem appears to be endemic
to a large number of studies from the past several decades, as discussed in more detail in Section \ref{sec:conclusions}.

This, and other related age-dating problems, are investigated in the current paper. We develop methods which improve the age-dating results for 10 to 20 \% of the overall cluster populations in four PHANGS-HST spiral galaxies. While this is a relatively small improvement in terms of the overall percentage of affected clusters, the improvements are quite important for studies of old globular clusters, and  can also impact a variety of important diagnostics for younger clusters since most of the old clusters  are {\it systematically} redistributed to have ages around either 5 or 100 Myr. Examples of science results that may be affected by this problem are:
%studies which are affected  by this problem are: 
the power-law index of cluster age distributions (e.g., see Section \ref{sec:age_and_mass}), the presence or absences of a downturn at the upper end of the cluster mass function (e.g., \citealp{adamo17}, \citealp{mok19}), the fraction of stars that form in clusters (i.e., $\Gamma$ - \citealp{bastian08}, \citealp{kruijssen12}, \citealp{chandar17}), and the specific frequency of old globular clusters in spiral galaxies \citep{harris91}.  
We examine  the impact on the age distribution in this paper
% and $\Gamma$ 
for one galaxy, NGC~628, as an example.

\begin{figure*}
\begin{center}
\includegraphics[width =7.0in , angle= 0]{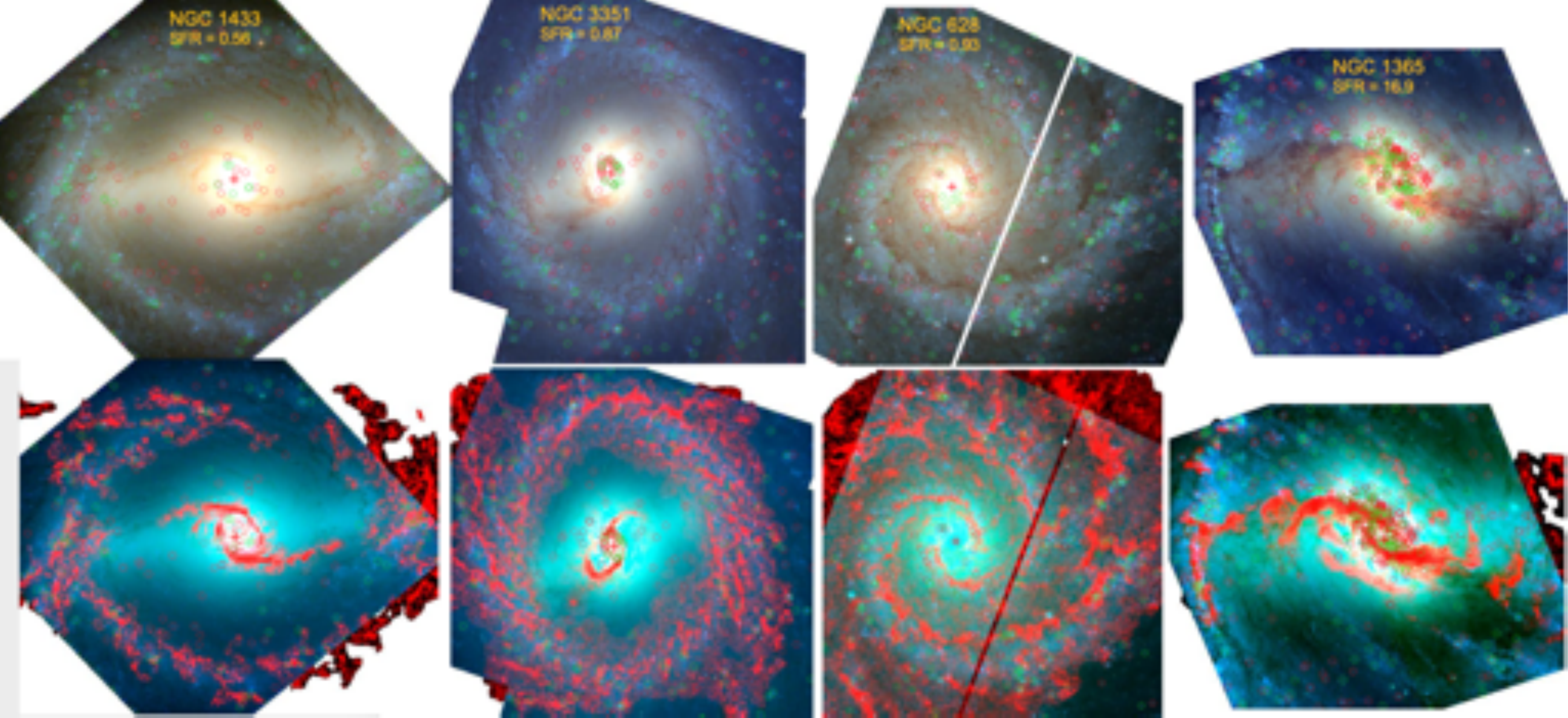}
\end{center}
\caption{
Hubble B-V-I images (upper: i.e., blue = F435W, green = F555W, and red = F814W) and B-V-CO images (lower: i.e., blue = F435W, green = F555W, and red = CO from ALMA) of the four program galaxies. The galaxies are ordered from low star formation rate (left) to high star formation rate  galaxies (right). Red circles are class 1 clusters (symmetric) while green circles are class 2 (asymmetric) clusters \citep{whitmore21}.  }
\label{fig:4x2_image}
\end{figure*}

 The current paper is  a pilot project using four PHANGS-HST galaxies (NGC~628, NGC~1365, NGC~1433, and NGC~3351, as shown in Figure \ref{fig:4x2_image}). The project is  designed to identify and fix the most common problems  in the determination of cluster ages, masses, and reddening as performed in the PHANGS-HST data reduction pipeline (i.e., \citealp{turner21}). 
The 
development of more sophisticated methods to optimize  age estimates for the full sample will be briefly discussed, and its incorporation into the PHANGS-HST pipeline will be described in a future paper. 

Our approach in this paper is to first identify a set of clusters with age estimates which are clearly incorrect. We then experiment with how to objectively identify them in a color-color diagram, and establish rules that fix the ages for most of these clusters. The few remaining exceptions (e.g., primarily young clusters with significant reddening due to dust that gives them colors similar old globular clusters) are then corrected using information from either H$\alpha$, CO intensity, or dust morphology. We find that, perhaps not surprisingly, H$\alpha$ emission offers the most direct and powerful way to establish the presence of young stars, but in the absence of such data, tracers that indicate the presence of dust,
%the environmental abundance of dust, 
such as CO or dust morphology, can also be used.

The rest of this paper is organized as follows. Section
\ref{sec:data} 
describes the data and sample.  In Section \ref{sec:problems} we present and discuss the primary age-dating problems we want to solve.
% \ref{sec:procedure}  
In Section \ref{sec:approach}  
we describe the procedure used to identify "bad ages" and investigate where they are generally found in the log Age vs. $E(B-V)$ diagram and the U-B vs. V-I diagram. In Section \ref{sec:solutions} we describe how we build the hybrid age-solutions for clusters in the four program galaxies, starting with the easiest galaxies  (NGC~1433 - low Star Formation Rate, hereafter SFR, and relatively minor reddening) and finishing with the most challenging (NGC~1365 - high SFR and extensive dust). 
%% Section 
% \ref{sec:decision_tree} \ref{sec:solutions} collects what was learned in the previous section into a decision tree used for the age-dating procedure . 
Section 
%  \ref{sec:whats_best} 
\ref{sec:halpha_co_dust} 
examines how well H$\alpha$ flux, CO intensity, and dust morphology work to 
%three different observational parameters that can be used to help 
identify exceptions to our basic rules (i.e. "interlopers").
%, i.e.,  H$\alpha$ flux, CO intensity, and dust morphology.  
Section 
% \ref{sec:dndt}
\ref{sec:age_and_mass}
examines the impact of the new age-dating solutions on science results, in particular on the age distribution and fraction of stars found in clusters for NGC 628. Section \ref{sec:future_work} describes future work while Section \ref{sec:conclusions} provides a  summary and conclusions. 

\bigskip

%\subsection{Background - Sample}
% \section{Cluster Catalog and Other Datasets}
\section{Sample and Related  Datasets}
\label{sec:data}

\subsection{Pilot Galaxy Sample}\label{sec:sample}
Four galaxies from the PHANGS-HST survey \citep{lee22} have been chosen for this exploratory study. These include galaxies that span the range from low SFR and low dust content, to high SFR and dust content.
% as shown in Figure \ref{fig:4x2_image}. 
The galaxies are
NGC 1433 (a relatively dust-free, low SFR barred spiral), NGC 3351 (a barred spiral with low dust content and SFR in the outer region, but a high SFR ring with extensive dust in the inner region), NGC 628 (a grand-design spiral with moderate SFR and dust), and NGC 1365 (a dusty barred galaxy with very high SFR). Galaxy distances range from  10 Mpc (NGC 628 and NGC 3351), to 18 Mpc for NGC 1433, and 20 Mpc for NGC 1365 \citep{gagandeep21}. 

The four galaxies are shown in Figure \ref{fig:4x2_image} in the order they will be discussed in this paper, from the lowest SFR (0.56 M$_\odot$/yr), least dusty galaxy (NGC 1433) to the highest SFR (16.90 M$_\odot$/yr) dustiest galaxy (NGC~1365). The top panels show the F814W, F555W, F438W composite color image using software adopted from the Hubble Legacy Archive \citep{whitmore16}, while the bottom panels show versions with the ALMA CO (2-1) emission in red \citep{leroy21}. The red circles are class 1 (symmetric, centrally peaked: hereafter C1) clusters, while the green circles show the class 2 (asymmetric, centrally peaked: hereafter C2) clusters, as discussed in \cite{whitmore21}. These will be the only two classes included in the current study (i.e., class 3 and multi-scale associations - \citealp{larson22},  are not included).

\subsection{HST Observations}
\label{sec:hst_data}

The primary datasets used in this paper are from the PHANGS-HST project (PI: J.C. Lee, GO-15654)
% Physics at High Angular Resolution in Nearby Galaxies (PHANGS) project taken with the Hubble Space Telescope (PHANGS-HST; PI: J.C. Lee, GO-15654)
\citealt{lee22}.\footnote{\url{https://archive.stsci.edu/hlsp/phangs-hst}}  PHANGS-HST is a Cycle 26 Treasury program which obtained multi-band imaging for 38 nearby spiral galaxies in the following five filters: F275W (NUV), F336W (U), F438W (B), F555W (V), F814W (I), with the WFC3 camera (or ACS in some cases with existing data).  

%Summarize the basics of the cluster catalogs:
The PHANGS-HST  catalogs of compact star clusters are one of the key data products from the project.  Cluster candidates are  selected using a Multi-Concentration-Index (MCI), as described in \citet{thilker22}. They are then classified into four classes based on visual examination (see \citealp{whitmore21}). 
In this work, we use only the human classified, C1 
% (symmetric, centrally peaked) clusters 
and C2 
%(asymmetric, centrally peaked)
clusters. These human classified clusters have also been used to train a convolutional neural network to obtain machine learning classifications for a larger (fainter) sample of star clusters (see \citealp{wei20},   \citealp{whitmore21}, and Hannon et al. 2023 - in preparation). 
%% Only the human classified clusters are used in the current paper.  
The numbers of  objects for each target galaxy are compiled in Table \ref{tab:table1} and range from N = 191 for NGC~1433 to N = 635 for NGC~1365.  

%All PHANGS galaxies have CO(2-1) observations from the PHANGS-ALMA large program \citep{leroy21}. \footnote{\url{https://sites.google.com/view/phangs/home}}.
The age-dating results presented throughout this paper are based on SED fits between cluster photometry in the NUV, U, B, V, and I bands and the Bruzual \& Charlot (2003) stellar population models, through the CIGALE fitting software \citep{Boquien19}. For the original baseline  solution, 
we adopted a fixed solar metallicity ($Z=0.02$) model, and fit for the best combination of age and reddening, where the ages range from 
%%% log~$(\tau/\mbox{yr})=6.0-10.2$ 
log~Age = 6.0-10.2 yr,
and the reddening between $0.0 - 1.5$~mag. For the hybrid solutions that will be discussed later in this paper, we also performed fits with 1/50th solar metallicity ($Z=0.0004$), that are used for likely old globular clusters.
The \cite{fitzpatrick99} reddening law with 
R$_{V}$ = 3.1 is used. 
This has been shown to better match clusters in spiral galaxies when using deterministic predictions like the Bruzual \& Charlot models.  Cluster age and reddening estimates from SED fitting are not very sensitive to the adopted extinction law in general. For example, the estimated ages for only a handful of clusters change by $\sim \pm0.1$ in log~age when the Calzetti attenuation curve is adopted instead.

As described in \citet{turner21}, about 20 \% of the clusters 
% in NGC 3351 
have bimodal probability distributions indicative of the age/reddening degeneracy problem discussed in the current paper. If these objects are excluded, the average uncertainty in our cluster age estimates is about 0.3 dex. An example of a bimodal probability distribution is shown in Section \ref{sec:common_problem_2}.

Age estimates from Version 1.1 of the PHANGS-HST pipeline are used in this study. 
The primary difference from Version 1, which is  described in \citet{turner21}, is the use of developmental software to include upper limits for the fluxes in cases where the values are less than $1\sigma$ of the photometric error estimate, as described in more detail in Deger et al. (2023 - in preparation). This is primarily relevant for the NUV (F275W) and U (F336W) photometry, and affects only a few percent of the class 1 and class 2 clusters.  Catalogs developed as part of  the current paper will be published in the PHANGS-HST archives and new updated catalogs will be made available as the PHANGS-HST pipeline evolves.  
Archival H$\alpha$ imaging from HST for NGC~628, NGC~1433, and NGC~3351 were also examined and used to make snapshots to illustrate various points in this paper. Unfortunately, no HST H$\alpha$ imaging is available for NGC~1365. 

PHANGS-HST is creating new dust extinction maps (as described in \citealp{thilker23}), which we use in the current paper.
Dust lanes are identified as dark, filamentary structures in each galaxy using a machine learning approach (specifically: U-Net), operating on the F336W, F438W, and F555W images. At the time of submission, 
%of the current paper the Thilker project 
the project had not yet generated extinction maps ($A_V$ versus position). Instead, we use 
masks which identify the dust features (0 or 1, depending on the presence of a dust lane) from early versions of the  project.
These are discussed  in Section \ref{sec:dust_halpha}, where we use them to test how well they identify interlopers.  

\subsection{MUSE and ALMA Observations}
\label{sec:muse_alma_obs}
 Since high-resolution, H$\alpha$ images with HST are not currently available for most of the galaxies in the PHANGS-HST sample, H$\alpha$
  measurements at the locations of all cluster candidates from PHANGS-MUSE (for 19 of the 38 PHANGS-HST galaxies) are used in this paper. While this provides a uniform dataset, it has the disadvantages of lower resolution (approximately 0.8$\arcsec$ for MUSE, compared to $0.1\arcsec$ for HST). The lower resolution, for both the MUSE H$\alpha$ 
  % (approximately 0.8 arcsec) 
  and the CO observations ($\approx1.2\arcsec$), is one of the primary limitations of our pilot study. 
  The H$\alpha$ flux measurements are determined from the MUSE images using aperture photometry with a radius that matches the MUSE resolution. No continuum or annular sky subtraction is performed.  
  % of 0.8 arcsecand 
%  No continuum subtraction is used.
  % or sky subtraction. 
  See \citet{emsellem22} for further details about the MUSE project.

%\subsection{ALMA Observations}

Similar integrated CO(2-1) line intensity measurements at the locations of the cluster candidates were made using the ALMA  observations, as described in \citet{leroy21}.
Given that the ALMA resolution element is much larger than the circular aperture used for HST photometry, we simply extract the CO intensity on the ALMA map at the pixel nearest to the cluster sky position.   For a typical PHANGS-HST galaxy at 16 Mpc median distance, the 1.2 arcsec ALMA beam size corresponds to  approximately 90 pc.
%%% compared to a full well half maximum (FWHM) value of 0.1" for the HST observation corresponding to a value of approximately 10 pc. 
Hence,  the CO intensity at the positions of the clusters is often under or over estimated due to smoothing effects. This will be discussed in more detail in Section \ref{sec:best-interloper}.

%\section{Background: Data, Sample, and Primary Age-dating Problems}
\section{The Primary Problems When Age-dating Star Cluster Populations}
\label{sec:problems}

\begin{figure}
\begin{center}
\includegraphics[width =3.3in , angle= 0]{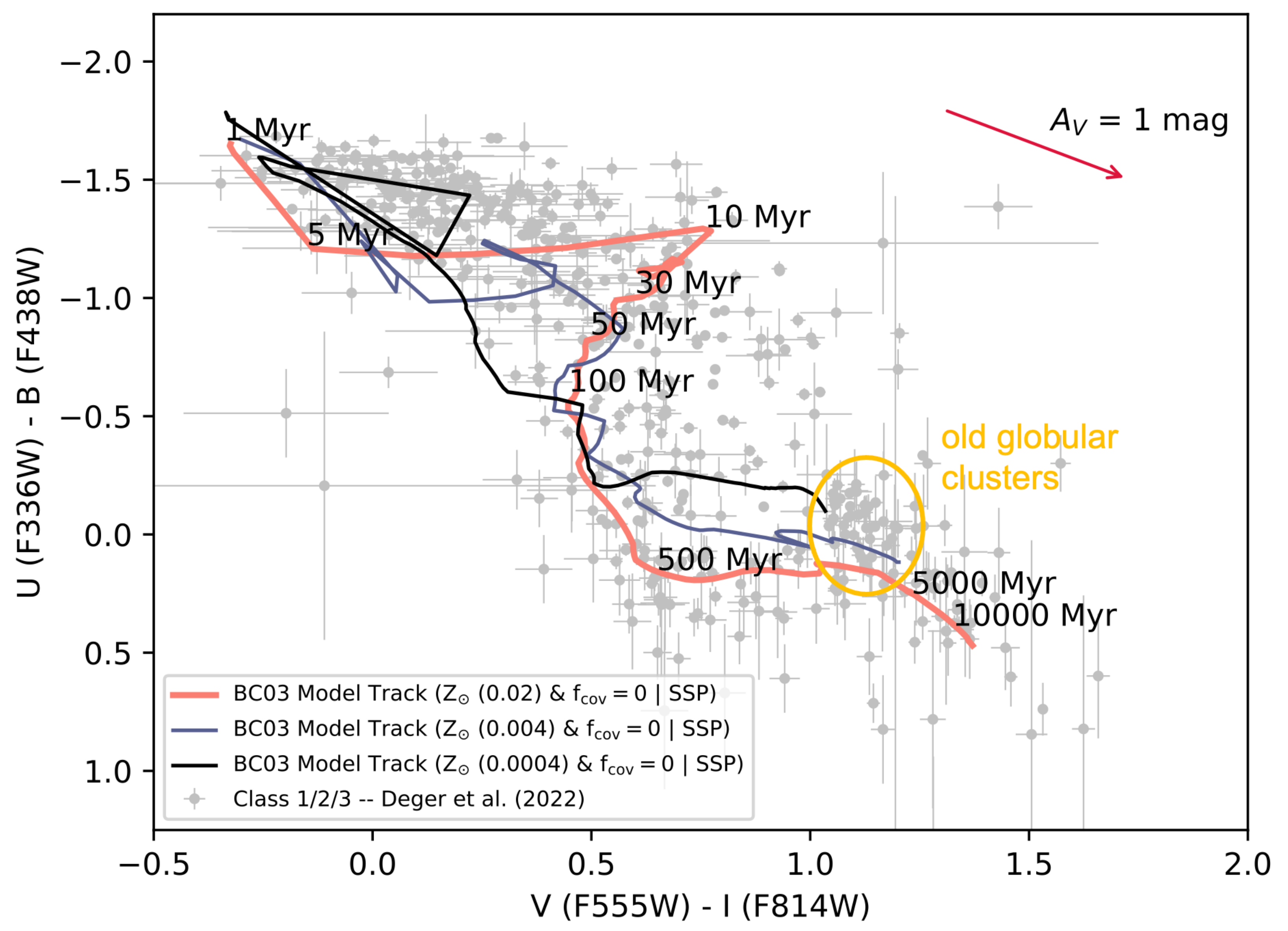}

\end{center}
\caption{$U-B$ vs. $V-I$ color-color diagram for bright, relatively isolated star clusters in  17 PHANGS galaxies  (see \citealp{deger22} for details). Three different SED tracks are shown: solar (Z$\odot$ = 0.02), 1/5th solar (Z$\odot$ = 0.004), and 1/50th solar (Z$\odot$ = 0.0004).
Ages for the solar metallicity track are included. The location of a high concentration of old globular clusters is shown by the yellow oval, which falls just below  the end of the 1/50th metallicity track (i.e., with small amounts of reddening). The reddening vector is shown by the red arrow.
}
\label{fig:cc_3metal}
\end{figure}
 
%\subsection{The Age-Metallicity Degeneracy and Old Globular Clusters}
\subsection{Problem 1: Star clusters do not all have the same metallicity}
\label{sec:common_problem_1}

%age-metallicity degeneracy
Most studies that perform SED age-dating for stellar clusters assume a single metallicity, generally solar or near solar for spirals, 
% and more massive galaxies, 
and subsolar for dwarfs and irregulars. This is a natural choice since the focus is generally on recently formed clusters, which have higher metallicity than ancient clusters.
If old globular clusters are included in the fitting, there can be a mismatch between some of their observed colors and the higher metallicity predictions,
%observations and the model being used to predict their colors, 
which can result in incorrect
%large errors in estimates o
age and reddening determinations.
These incorrect results for ancient clusters pollute the young and intermediate age populations.
The broad wavelength coverage of the PHANGS-HST survey makes our dataset well-suited for studying both young and old clusters. In order to provide accurate estimates for clusters of all ages, 
we need to include fits to both high and low metallicity models in our age-dating procedure.

Figure \ref{fig:cc_3metal} illustrates the age-metallicity degeneracy by showing the measured colors of bright, isolated clusters (used to determine aperture corrections) in 17 PHANGS-HST galaxies from
\citet{deger22}. 
These  are often old globular clusters in the outskirts of galaxies. 
The clump of points around $V-I = 1.1$, $U-B = -0.2$ in Figure \ref{fig:cc_3metal} (i.e., within the yellow oval) are nearly all old globular clusters, based on visual examination (they are yellowish in color, generally isolated, and show no evidence of dust or H$\alpha$ emission).
Note how far away they are from the the end of the orange (solar metallicity) SED model at $V-I = 1.4$, $U-B = 0.5$ (that they are being fit to in the PHANGS age-dating pipeline), and how close they are to the 
%Instead, they are very near the 
end of the 1/50th solar metallicity shown by the green line.

Redder, more metal-rich, globular cluster (for example those associated with bulges) are also probably present, but fall along the solar metallicity BC03 model predictions, and hence are much more likely to be age-dated correctly. While second parameter effects, such as horizontal branch structure, may also have some impact on the colors of globular clusters, these do not impact our primary goal of identifying old globular clusters and developing methods that allow us to approximately age-date them correctly.
% (i.e., 0.0004), 
%as would be expected for old globular clusters. 

In what follows, we will assume solar metallicity for the young clusters and 1/50th metallicity for objects we believe to be old globular clusters. In principle, we could attempt to use different metallicities for different galaxies, e.g., using the values in \citet{santoro22}.
However, the grid of available metallicity models is fairly coarse, and nearly all of the galaxies in our sample would use the solar metallicity model in any case. Similarly, since there is a gradient in the metallicities as a function of radius for many spiral galaxies (e.g., \citealp{sanchez14},
\citealp{kreckel19}, \citealp{williams22}), we could attempt to include this dependency. However, the gradient is generally weak, and represents a much smaller issue than the age/reddening/metallicity degeneracy which is the main focus of this paper. We plan to make a more detailed study, with more  finely sampled metallicity values, as part of the future work discussed in Section \ref{sec:future_work}.

The metallicities for old globular clusters in our galaxies are essentially unknown, hence we will use the relatively good agreement between the clump of points in Figure \ref{fig:cc_3metal} and the bottom end of the 1/50th metallicity track as justification for using the same value for all of our galaxies.  This is a typical value for normal spiral galaxies. For example, \citet{lomel22} 
use a range of 1/20 to 1/50 solar metallicity to fit their sample of old globular clusters in spiral galaxies. Several studies find that changing metallicities by only a factor of a few has relatively minor effects on the results (e.g., \citealp{whitmore20}, \citealp{messa21}). It is primarily when changes on the order of ten or more are made, as in the current paper, that  large differences in age estimates are found.

%\subsection{Age-reddening degeneracy: ancient clusters mistaken for young}
\subsection{Problem 2: Allowing reddening to be a free parameter can result in large errors }

\label{sec:common_problem_2}

Most SED age-dating software constructs a grid of luminosity and color predictions spanning a range of age, reddening ($E(B-V)$) or extinction ($A_V$), and metallicity, then finds the combination of these parameters that  minimizes the difference between observed  and predicted fluxes in a number of photometric bands (generally four or more). In most recent studies, the metallicity is fixed while age and reddening are free parameters, albeit within some limits. 

If the assumed population synthesis model is not appropriate for the objects, for example adopting a solar metallicity model while trying to fit globular clusters with low metallicity, 
the cluster properties can be sufficiently different from the predictions that the program is unable to find the correct age (and reddening) solution.
In this specific case, the fitting program will often find an apparent solution "up the reddening vector" where a combination of age, and moderate to high reddening, gives a better match to the observed colors, rather than finding the correct age. 

\begin{figure*}
\begin{center}
\includegraphics[width =7.0in , angle= 0]{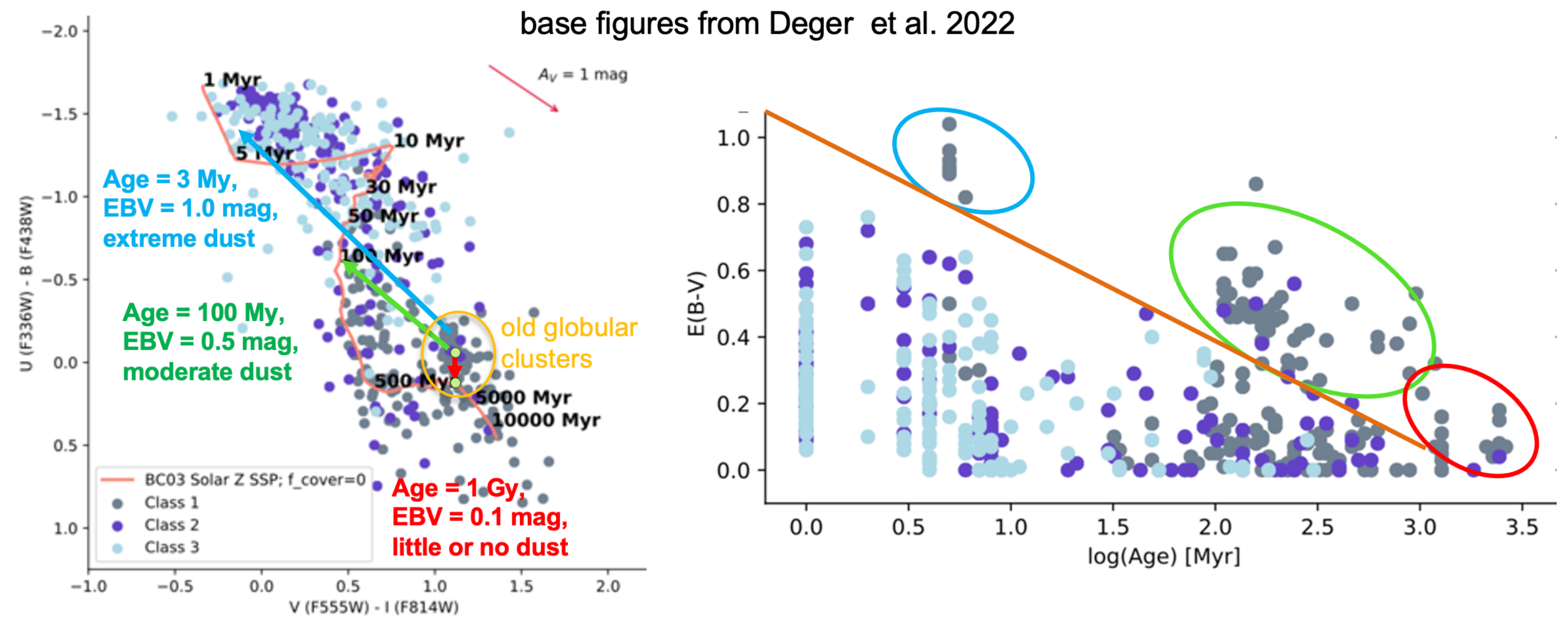}

\end{center}
\caption{Illustration of how the age/reddening/metallicity degeneracy (left panel) can lead to the incorrect age-dating of old globular clusters, resulting in positions above a diagonal line in the log Age - $E(B-V)$ diagram (right panel). Using figures from \citet{deger22} for 17 PHANGS galaxies as the base, the figure  shows how the mismatch between the location of the observations of old globular clusters (which have low metallicity)  and the pink SED track used for the prediction (solar metallicity) results in three different regions of poor fits, (i.e., $\approx$ 3 Myr shown by the blue arrow and oval, $\approx$ 100 Myr shown by the green arrow and oval, and $\approx$ 1 Gyr shown by the  red arrow and  oval), rather than the correct solution at the end of the SED track at around 10 Gyr (i.e., log Age = 4).  }
\label{fig:degen}
\end{figure*}

Figure \ref{fig:degen} demonstrates how the combination of a mismatch in metallicity, the inherent age-metallicity degeneracy, and the option to fit $E(B-V)$ reddening as a free parameter, can lead to 
many incorrect age solutions for old globular clusters. Using Figures 2 and 3 from \cite{deger22} as the base figures,  we again note the location of old globular clusters in the color-color diagram shown in Figure \ref{fig:cc_3metal}, with the yellow circle around them.
%clump of Class 1 (symmetric) clusters around $U-B = -0.2, V-I = 1.1$, as shown in Figure \ref{fig:cc_3metal} with the yellow circle around them. These are primarily old globular clusters.  

As graphically illustrated\footnote{It is important to note that most age-dating methods perform SED fits to age and reddening using all available photometric bands (e.g., F275W, F336W, F435W, F555W, F814W in our PHANGS-HST project),
rather than fitting in the 2-dimensional color-color diagrams shown throughout this paper. The two color diagrams, however, often make it easier to visualize what is happening in the fit, and hence are used extensively in the current paper. } in Figure \ref{fig:degen}, SED-fitting generally selects between three potential age-reddening combinations for these old globular clusters. The first solution (the red arrow originating from the position of the old globular clusters in the left panel and the red oval in the right panel) is a combination of old age (typically approximately 1 Gyr for solar metallicity) and very low reddening, where  the fit  effectively jumps down to the nearest position on the solar metallicity SED track.
%, i.e., the red arrow. 
%This results in ages around 1 Gyr if a solar metallicity track is being used, and very low reddening values. 
The resulting position in the log Age - $E(B-V)$ plot is shown by the red oval in the right panel of Figure \ref{fig:degen}. Note that estimated ages of $\approx1$~Gyr are still far from the correct answer of $\sim10$~Gyr.

In our pilot study, we find the most common solution is an intermediate age ($\approx100$ Myr) and reddening ($E(B-V)$ $\approx0.5$ mag), which is shown by the green arrow  and corresponding green oval  in Figure \ref{fig:degen}. This solution backtracks up the reddening vector until it hits the solar metallicity track at an intermediate age. This results in 
$E(B-V)$ values of $\approx$ 0.5 mag; clearly too high based on visual examination which show little dust around nearly all globular clusters. Roughly 60~\% of the old globular clusters in our sample are assigned incorrect ages in this intermediate range. 

The third possible solution (shown by the blue arrow and corresponding blue oval) is a combination of very young age (Age $\approx$ 3~Myr) and high reddening ($E(B-V)$ $\approx1$~mag). In this case, the fitting has backtracked all the way up the reddening vector,
to the youngest end of the SED track.
Though less common in our sample, this solution is still preferred by the fitting software between 10 and 20 \% of the time for old globular clusters. For example, for NGC 4449 the LEGUS age estimates based on a similar filter combination (\citet{calzetti15a}  assigns 2 of 6 spectroscopically-confirmed old globular clusters to this age range  \citep{whitmore20}.

The right panel in Figure \ref{fig:degen} shows that these three incorrect solutions populate a diagonal region in the log Age - $E(B-V)$ diagram above the rest of the cluster population, essentially showing the intersection of the 'backtracked' reddening vector and the SED track.

There are a number of other examples and recent discussions of this behavior in the literature,
for example \citet{turner21}, \citet{deger22}, \citet{hannon22} and
\citet{moeller22}.  We examine the scientific impacts that this type of incorrect age-dating can have by examining the case of NGC~628 in Section \ref{sec:age_and_mass}.

\begin{figure}
\begin{center}
\includegraphics[width =3.3in , angle= 0]{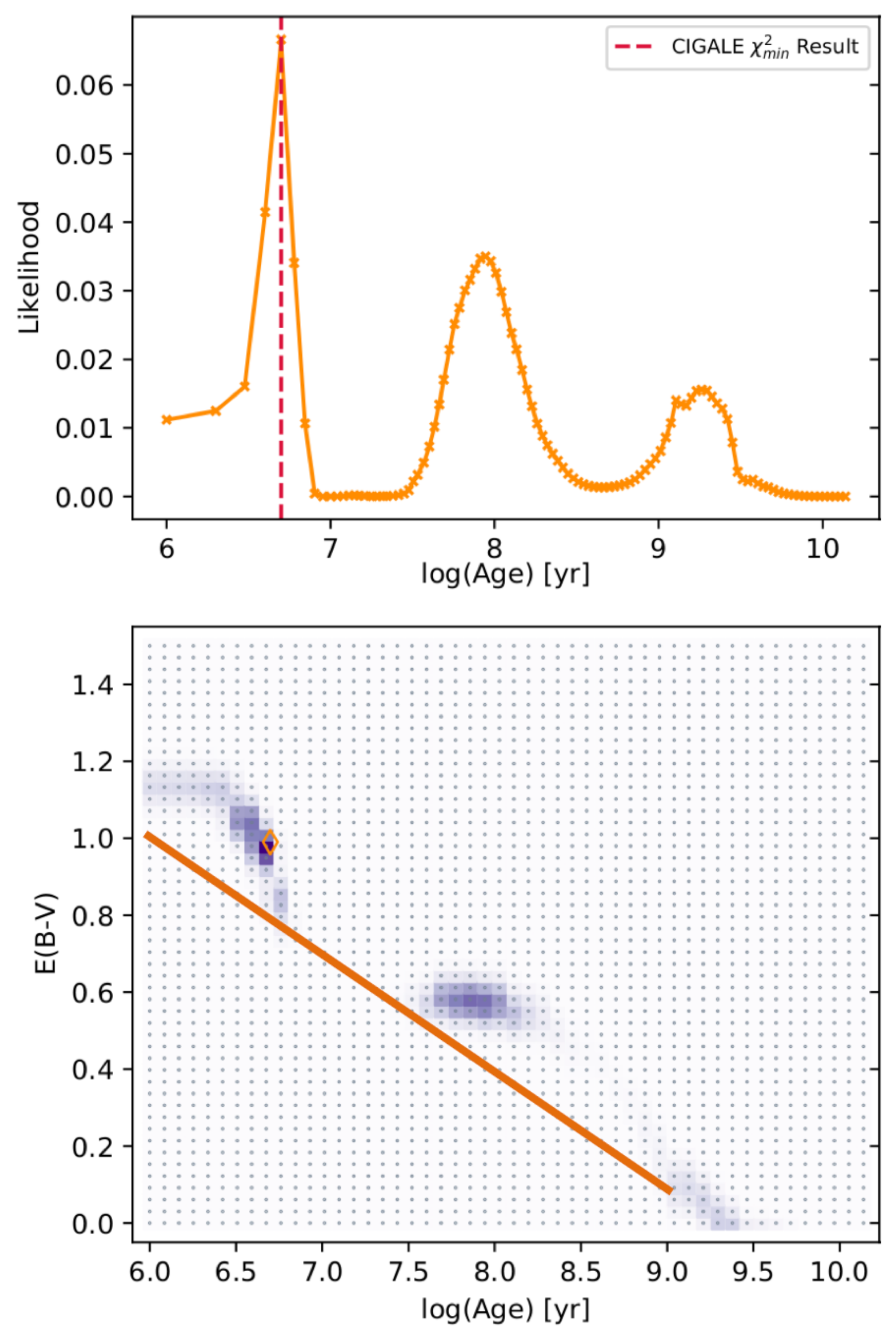}
\end{center}
\caption{Illustration of the multi-modal probability diagrams from the CIGALE results \citep{Boquien19}. This is for an old globular cluster (object \# 687 in NGC~1365 - see  snapshot in Figure \ref{fig:7plot-n1365}). Note how the mismatch between the low metallicity globular cluster and the solar metallicity SED used to fit it results in an incorrect largest likelihood peak being selected, predicting an age of 7 Myr for an old globular cluster. Also note the similarity of the bottom  diagram with the right panel of Figure \ref{fig:degen}, explaining why the bad ages end up above the same diagonal line. See Deger et al. (2023 - in preparation) for more details about this figure.}
\label{fig:bimodal}
\end{figure}

Another useful tool for visualizing the three possible solutions 
%for objects which are likely to be globular clusters 
is a probability distribution plot, as derived from the CIGALE output. These plots will be one of the primary tools used in the followup to the current pilot study, as discussed in Section \ref{sec:future_work} (see also \citealp{turner21}).
An example of a probability distribution plot is shown in Figure \ref{fig:bimodal} for  object \# 687, a candidate globular cluster in NGC 1365 (see discussion and snapshot in  Section \ref{sec:solutions}). The three peaks correspond to the three potential age-reddening solutions shown in Figure \ref{fig:degen}.

The top plot in Figure \ref{fig:bimodal} shows that the strongest likelihood peak is the youngest one, at log Age = 6.8 (i.e., 7 Myr), and that is what is reported, incorrectly, as the minimum $\chi^{2}$ result by CIGALE. However, the best solution, since visual inspection shows that these are generally old globular clusters given the absence of gas or dust in their vicinity,  is actually the  oldest solution at log Age = 9.4. The lower value of the likelihood  in the top panel of Figure \ref{fig:bimodal} for this older solution
is due to the mismatch in metallicity between old globular clusters and the solar metallicity being used for the fit (i.e., Problem \#1). The bottom plot in  Figure \ref{fig:bimodal} again demonstrates how old globular clusters with bad age estimates result in points above a diagonal in a log Age - $E(B-V)$ diagram. Notice the similarity with the right panel of Figure \ref{fig:degen}.

%\subsection{Age-reddening degeneracy: reddened young clusters masquerading as old}
\subsection{Problem 3: Some heavily reddened clusters require higher $E(B-V)$ limits}
\label{sec:common_problem_3}

For most galaxies in the PHANGS-HST and LEGUS samples, assuming a maximum reddening of around $E(B-V)$ = 1.5 is appropriate, since these systems contain a small to moderate amount of dust.  This assumption works reasonably well for three (NGC 628, NGC 1433, NGC 3351) of the four galaxies studied in this work.
However, for a few galaxies,  the spiral arms and central regions can be sufficiently dusty that a higher limit is needed.
%% to avoid overestimating the ages of very young, reddened clusters.
%, because otherwise they cannot reach the positions for a very young (e.g., 5 My) cluster. 
This particular issue will be examined for NGC~1365 in Section \ref{sec:hybrid-1365}.

This case results in  the opposite issue from problems 1 and 2:
the ages of young, dust-embedded clusters can be overestimated by factors of $\sim1000$, as also seen in  \cite{hannon22}.
% instead of old globular clusters being assigned ages that are too young, by factors of up to 1000.
%, the ages of young dust-embedded clusters can be overestimated by factors of 1000.
This third problem will become more relevant in the future, when JWST will allow researchers to study more heavily reddened and embedded clusters.

\section{Identifying Incorrect Age Results}
%\section{Procedure}
\label{sec:approach}

%\subsection{Visual examination to produce bad-age files}
\subsection{Visual Examination to Identify Bad Ages}
\label{sec:bad_age}

A brief look at the bulges of many spiral galaxies immediately reveals the clear presence of a population of old globular clusters. The clusters are uniformly yellow, isolated, and there are no blue objects nearby. %Examples in our sample are 
NGC~1433 and NGC~3351 provide particularly good examples of bulge globular clusters. These regions provide an initial training set of the properties of old globular clusters and how they appear in the HST images.

%Many of the problems with age-dating star clusters discussed in the current paper have been identified in previous papers, including \citet{whitmore20}.  \citet{turner21}) and \citet{deger21}. Based on these results, 
Based on these initial training sets, our first step 
% to correcting the age-dating issues discussed in the previous section 
is to perform a manual, systematic  search for  clusters with "bad ages" in the four PHANGS-HST galaxies studied here.
 This search was performed by co-author M. Floyd, who visually inspected color images (to check for the presence of dust that would be consistent with large $E(B-V)$ values)),  and inspected H$\alpha$  and ALMA CO (2-1) images (to check for the presence of ionized and molecular gas that would be consistent with stellar populations younger than approximately 10~Myr). He also compared the measured $B-V$ vs. $V-I$ colors with BC03 model predictions, and  examined the environment and isolation of each cluster. The visual inspections were performed  using the Hubble Legacy Archive and custom software \citep{whitmore16}. This part of the project is discussed in more detail in 
 Floyd et al. 2023 (in prep).
% An examination of the ALMA CO(1 -2) image was also made.

A designation of  "bad age" requires a difference  of a factor of ten or more between the visually estimated age and the version 1.0 SED age estimate, as described in \citet{turner21}. The most common age-dating mistake, as expected, are objects that are most likely old globular clusters (e.g., yellowish, no emission-line flux, no evidence of dust, often in the bulge or towards the outer portion of the galaxy, generally with $V-I > 0.9$ and $U-B > -0.5$), but are best fit by ages $\lea 500$ Myr and reddening $E(B-V)$$\gea0.4$~mag.

A similar examination was made in NGC~1433 by \cite{hannon22}, and eight clusters with underestimated ages were identified.
%allowed for an independent check of "bad-age" objects.
%, who found examples of underestimated ages for eight clusters in NGC 1433 for high mass clusters. 
Seven out of the same eight  clusters are also identified as having bad ages in the current work, while the eighth source has a reasonable best fit age of 
%in common with the current sample, with one cluster differing since PHANGS-HST gave it an old age of 
4 Gyr in the PHANGS-HST fitting procedure (i.e., it does not qualify as having a bad age) while the LEGUS pipeline assigned a best fit age of 3 Myr for this object, and hence it was included in the \citet{hannon19} review. 
Similar examples of underestimated ages for old globular clusters 
%(in many cases based on spectroscopic observations) 
are reported and discussed in detail for NGC~4449 in \cite{whitmore20}.

\begin{figure}
\begin{center}
\includegraphics[width =3.3in , angle= 0]{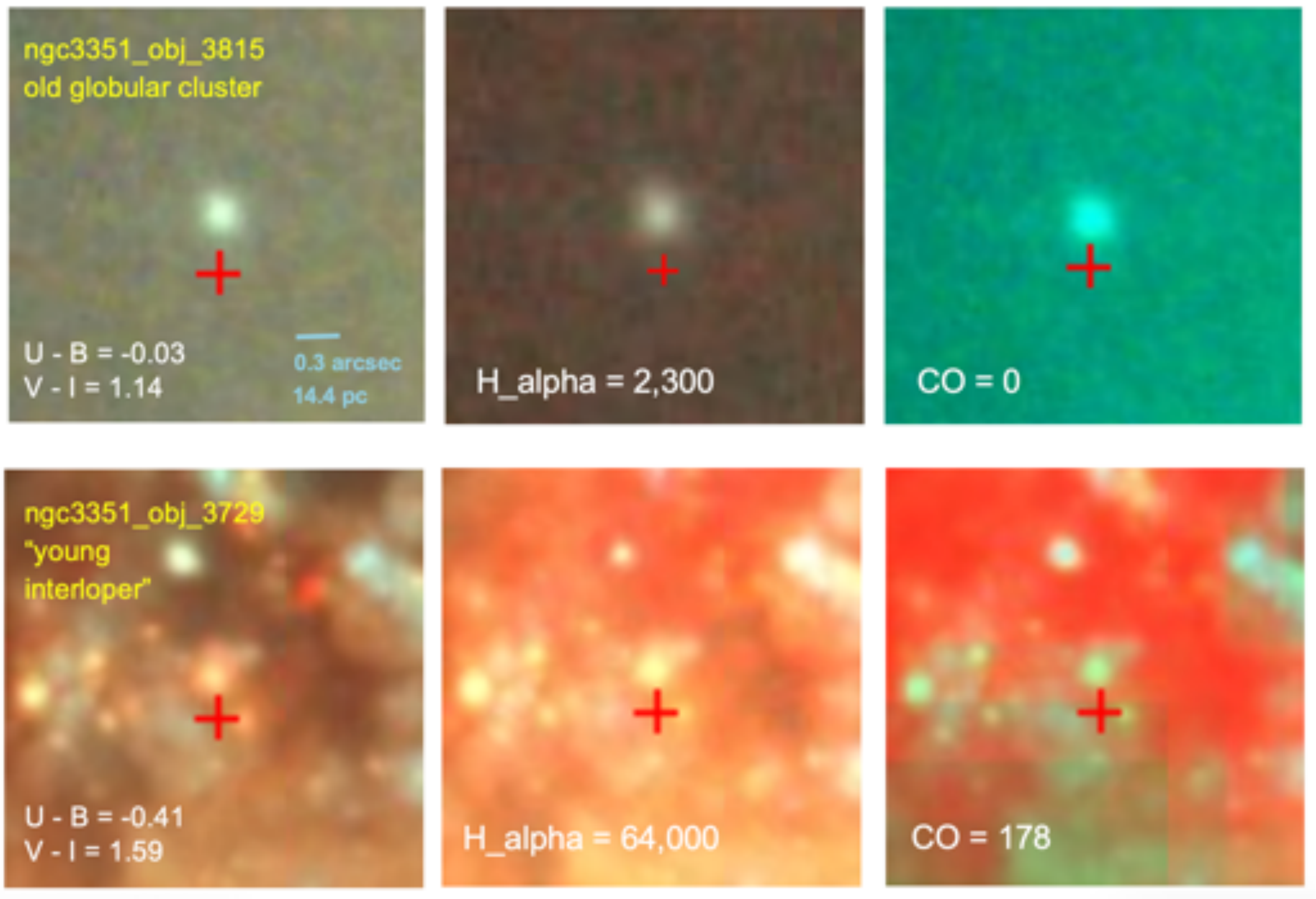}
\end{center}
\caption{Snapshot images of a typical old globular cluster in the top panels, and a typical "young interloper" (i.e., young dusty object with the same colors as an old globular cluster) in the bottom panels. The object of interest is always just above the red cross. The left panels are B-V-I images from Hubble; the center panels are B-V-H$\alpha$ images with H$\alpha$ in red; the right panels are B-V-CO images with CO from ALMA in red. The values for the H$\alpha$ flux (in units of $10^{-20}$ ergs/s/cm$^2$/pixel) and CO intensities (in units of K km s$^{-1}$) are also given. The blue bar in the upper left panel shows the scale. }
\label{fig:poster}
\end{figure}

Figure~\ref{fig:poster} shows an example of an old globular cluster (top panels)  and a "young interloper" (bottom panels - i.e. a young cluster with enough reddening from dust to give it  colors similar to an old globular cluster). The three columns 
show an optical Hubble B-V-I color image (left), 
a B-V-H$\alpha$ map (middle), and a B-V-CO (right) image.
The old globular cluster is intrinsically yellow and has very little dust or CO associated with it. In addition, 
the old globular cluster  is  quite isolated, and in a region with a smooth background, i.e., the bulge of NGC 3351. 

The young interloper has extensive H$\alpha$ around it, hence it is young ($<$ 10 Myr) and intrinsically blue, but has enough dust around it to give it $U-B$ and $V-I $ colors that are similar to a globular cluster, as listed  in the figure.   In addition, the young interloper  has many other sources  near it, some with blue colors indicative of young stars, as well as extensive CO, another indicator of a young age.

\subsection{Mapping Bad Ages to the log Age - $E(B-V)$ and Color-Color Diagrams}
%\subsection{Mapping the bad age objects to the log Age - $E(B-V)$ diagram and the U-B vs V-I color-color diagram}.

\begin{figure*}
\begin{center}
\includegraphics[width =7in , angle= 0]{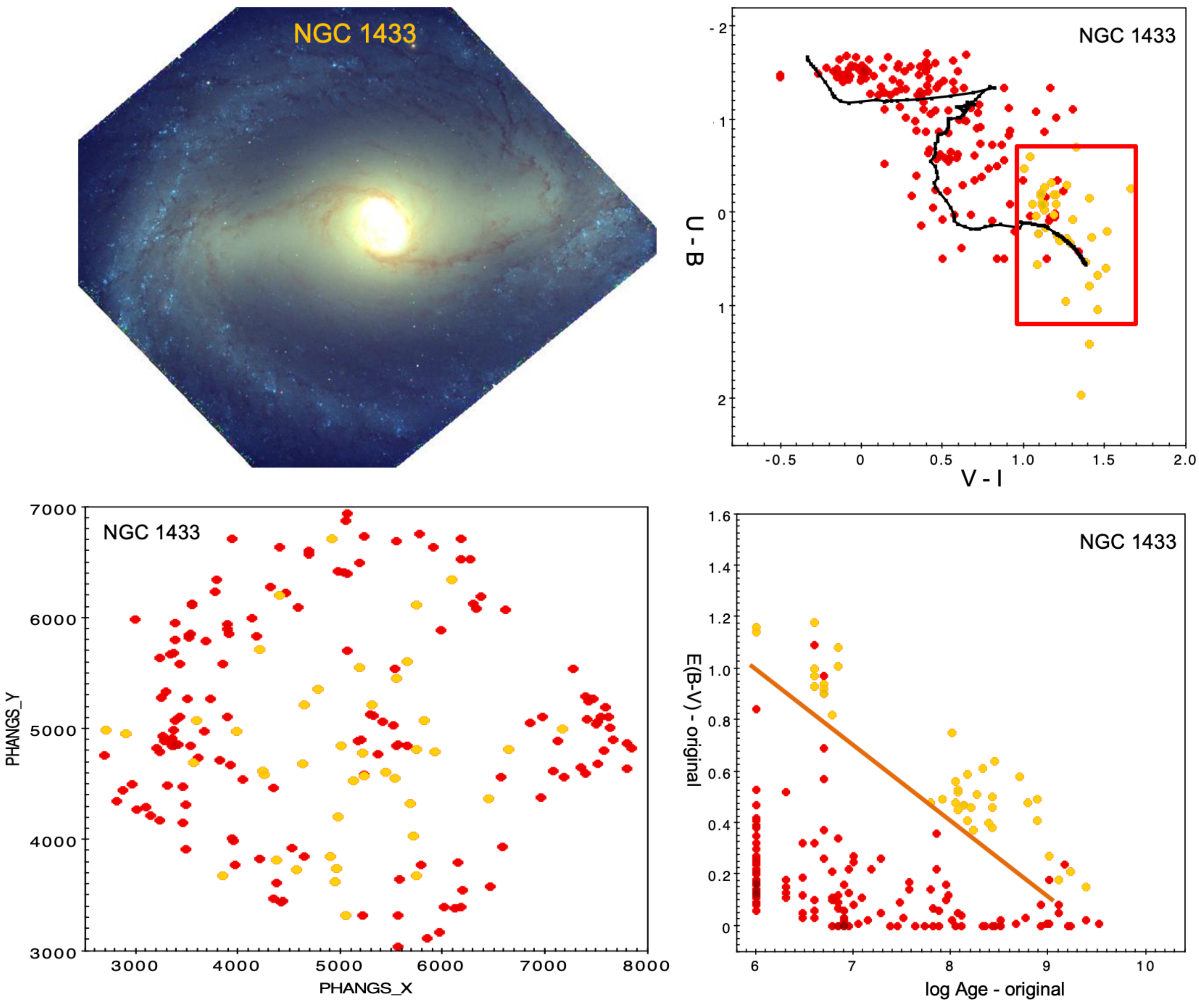}

\end{center}
\caption{Location of "bad ages"  (gold points) in the original age-estimate for NGC 1433 star clusters, as determined from a manual review by coauthor M. Floyd (see Floyd et al. 2023 - in preparation). The top left shows a Hubble B-V-I image. The bottom left panel shows the locations of the clusters with bad ages, with a  concentration in the bulge region where no evidence of recent star formation is found. The upper right panel shows that nearly all bad age estimates are in the region of the $U-B$ vs. $V-I$ diagram expected for old globular clusters; and the bottom right plot shows that nearly all the bad ages are found above a diagonal line in the log Age  vs. $E(B-V)$ diagram that runs from (6, 1.0) to (9.0, 0.1). The diagonal line used throughout this paper was originally defined based on this figure; i.e., it was placed just below where all the bad ages in NGC~1433 were found. }
\label{fig:n1433_cc_bad12}
\end{figure*}

Figure \ref{fig:n1433_cc_bad12} shows the locations of the bad-age objects (generally old globular clusters) identified in NGC~1433.
The upper left panel shows a B-V-I color image. 
The bottom-left panel shows the bad-age objects in gold. This highlights the fact that there are many bad age clusters in the bulge region.

The upper right panel  shows that in NGC~1433, nearly all of the bad-age objects are in the same region of the color-color diagram, highlighted by the red box.
This box covers the $U-B$ range from -0.6 to 1.2, and $V-I$ range from 0.95 to 1.7. The size of the box has been chosen to be large enough to contain all but two  of the clusters (with large uncertainties) that were determined to have bad ages. The strongest concentration of points is found just above the solar model predictions, as also demonstrated in Figure \ref{fig:cc_3metal} for a different sample, namely 17 galaxies that have been used to determine aperture corrections (see \citealp{deger22}). The box is somewhat wider than the concentration of lower-metallicity globular clusters in Figure \ref{fig:cc_3metal} to accomodate some of the lower S/N points (primarily due to larger uncertainties in the U-band), and the possibility of small amounts of reddening from dust. As will be demonstrated later in the paper, this works well for three of the four galaxies (NGC~1433, NGC~3351, and NGC~628), but will be modified somewhat in the case of NGC~1365 for a variety of reasons. 
We refer to this as the 'Old Globular Cluster' box (hereafter OGC-box) for the rest of the paper.  

The strong concentration of bad age points just above the solar model is the primary reason we are able to fix the age-metallicity problem for old globular clusters by adopting the 1/50th solar metallicity solution for clusters in this region of color-color space.
Most of the few remaining red points in the OGC-box  have age estimates around log Age = 9.5, and hence are not identified as bad ages since they are not a factor of 10 different from the typical age of an old globular cluster, which should have 
% log~$(\tau/\mbox{yr})\approx 10$. 
log Age $\approx 10$. 
None of the red points in the OGC-box for NGC 1433 in Figure \ref{fig:n1433_cc_bad12} are "young interlopers" (young, reddened objects which mimic the colors of old globular clusters). However, we will find that NGC 3351 and NGC 628 do have small numbers of young interlopers, as discussed in Sections \ref{sec:hybrid-3351} and \ref{sec:hybrid-628}.

\begin{figure*}
\begin{center}
\includegraphics[width =7in , angle= 0]{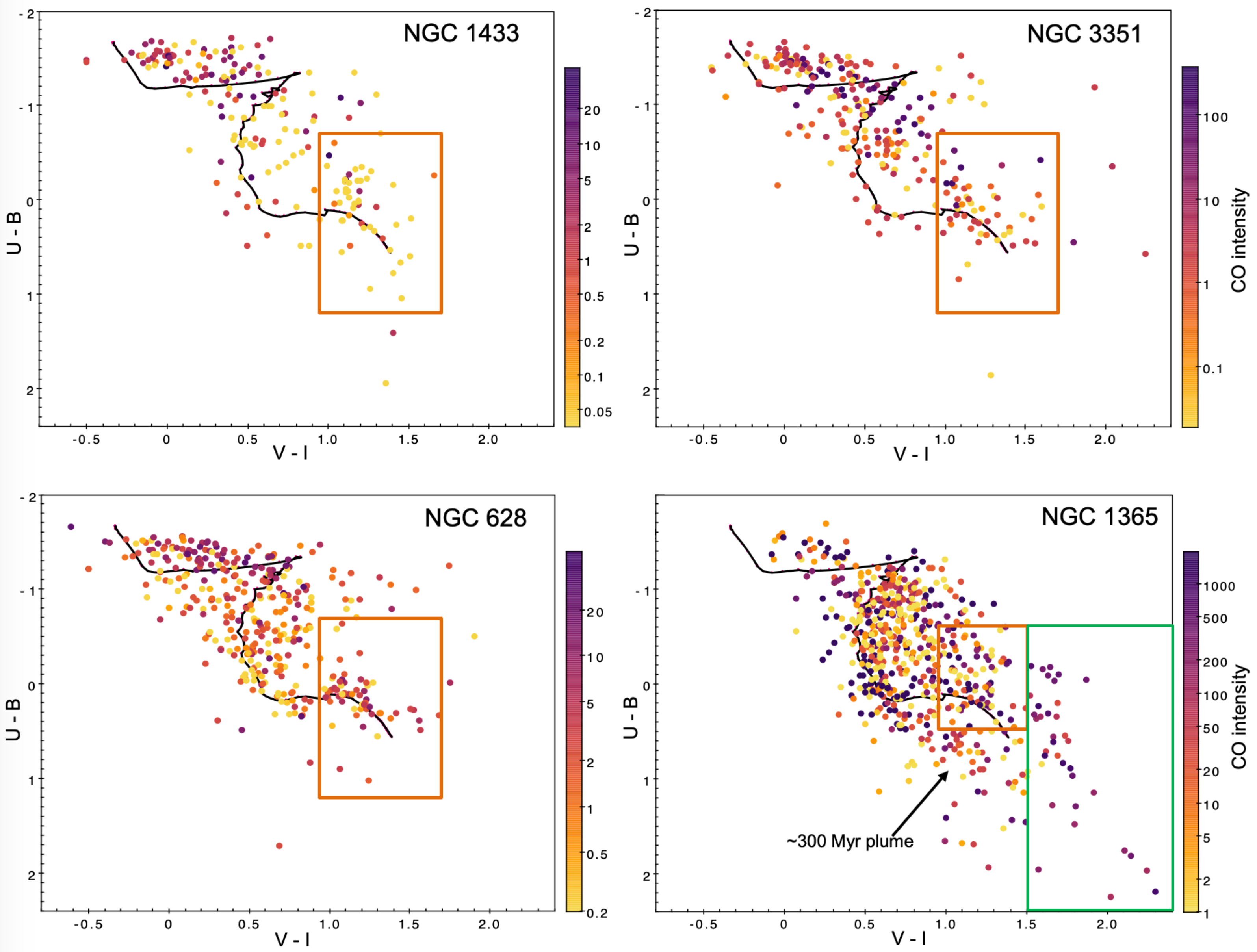}
\end{center}
\caption{Color-coded plot of CO intensities in $U-B$ vs. $V-I$  diagrams for the four program galaxies. A solar metallicity Bruzual-Charlot model is shown for comparison. The OGC box used to identify potential old globular clusters is shown by the orange box while the green box shows the  region used to identify the "high-red" objects in the case of NGC 1365 (see Section \ref{sec:solutions} for details). The 300 Myr plume objects in NGC 1365 discussed in Sections \ref{sec:hybrid-1365} and \ref{sec:300myr} are also identified.
CO intensity is in units of K km s$^{-1}$.
}
\label{fig:cc_4plot}
\end{figure*}

The bottom right plot shows an even more dramatic result, namely that all of the bad ages identified manually, and  completely independently, are above a diagonal line that runs from (log Age, $E(B-V)$) = (6, 1.0) to (9, 0.1).
This diagonal line 
%was originally defined based on this figure; i.e., it was placed just below where all the bad ages in NGC~1433 were found. However, subsequently we found the position of this line also worked quite well for all four of the galaxies in our sample, as discussed later in this paper. It also works for the multi-modal probability distributions shown in Figure  \ref{fig:bimodal}. Finally, it is 
works well not only for all four galaxies studied here but for the globular cluster populations in 18 PHANGS-HST galaxies studied by Floyd et al., 2023, and also for the multi-modal probability distributions shown in Figure  \ref{fig:bimodal}.
Hence, these two diagrams together provide an excellent way to identify clusters with potentially bad age estimates.

\subsection{$U-B$ vs. $V-I$ Diagrams as a Function of CO Flux }
\label{sec:cc_vs_co}

Figure \ref{fig:cc_4plot} shows the $U-B$ vs $V-I$ color-color diagram of clusters in all four target galaxies.  Here, the color coding shows the strength of the CO flux measured around each object. This provides a useful backdrop for the following discussion. 

We first note that NGC~1433
% the first galaxy we will examine, 
has very little dust and CO emission, with most of it associated with the youngest clusters in the upper left portion of the plot, as expected. This therefore provides a good starting point for our study since we expect reddening to have little affect on observed cluster colors. 
NGC~3351 and NGC~628 both show a bit more CO emission, some of which appears to be associated with a few clusters in the OGC-box (the dark circles in this box).
%the bottom right of the diagram. 

NGC~1365 looks dramatically different than the other three galaxies, both because of the very small number of young clusters in the upper left portion of the diagram, and also because 
a fair number of points are below or to the right of the OGC-box. Nearly all of these very red objects have strong CO (dark circles). This immediately shows that most of the objects in the bottom right part of the diagram are likely to be young, heavily reddened objects rather than old globular clusters. Because of this we will treat NGC~1365 differently than the other three galaxies.

% \subsection{Selection of Metallicity Solution and $E(B-V)$ Limit for the Hybrid Model in NGC 1433 }
% \label{age_ebv_3plot}

%\section{Building Hybrid models for Four PHANGS-HST Galaxies}
%\section{Using Lower Metallicity Models to Fix Incorrect Ages of Old Globular Clusters}
\section{Identifying and Fixing Incorrect Ages of Old Globular Clusters}
\label{sec:solutions}

Three out of four spiral galaxies studied here (and all but a few in the 38 PHANGS-HST sample) should have few intrinsically correct intermediate-age clusters with high reddening (i.e., above the diagonal line in a log Age vs $E(B-V)$ diagram), 
since they do not have much dust around them. When data points are seen above the diagonal, these are usually due to the incorrect age estimates resulting from the age/reddening degeneracy, with old globular clusters being mistaken for intermediate-age clusters with large reddening. In the next section we show how various hybrid solutions can be constructed to remove most of the age-dating errors and improve the success fractions by 10 to 20 \%, reaching values near 100 \% in most cases.

\subsection{Procedures}
\label{sec:procedure}

Here we present the techniques used to correct the age estimates (and associated estimates of reddening and  mass) that suffer from the problems described in Section \ref{sec:problems}. For most of the young clusters, the  original age estimates  are reasonably accurate (i.e., within 0.3 dex as discussed in Section \ref{sec:hst_data}), and we use the original solar metallicity fits from \cite{turner21}. However, for the likely old globular clusters we use a different set of rules.

\begin{figure*}
\begin{center}
\includegraphics[width =7in , angle= 0]{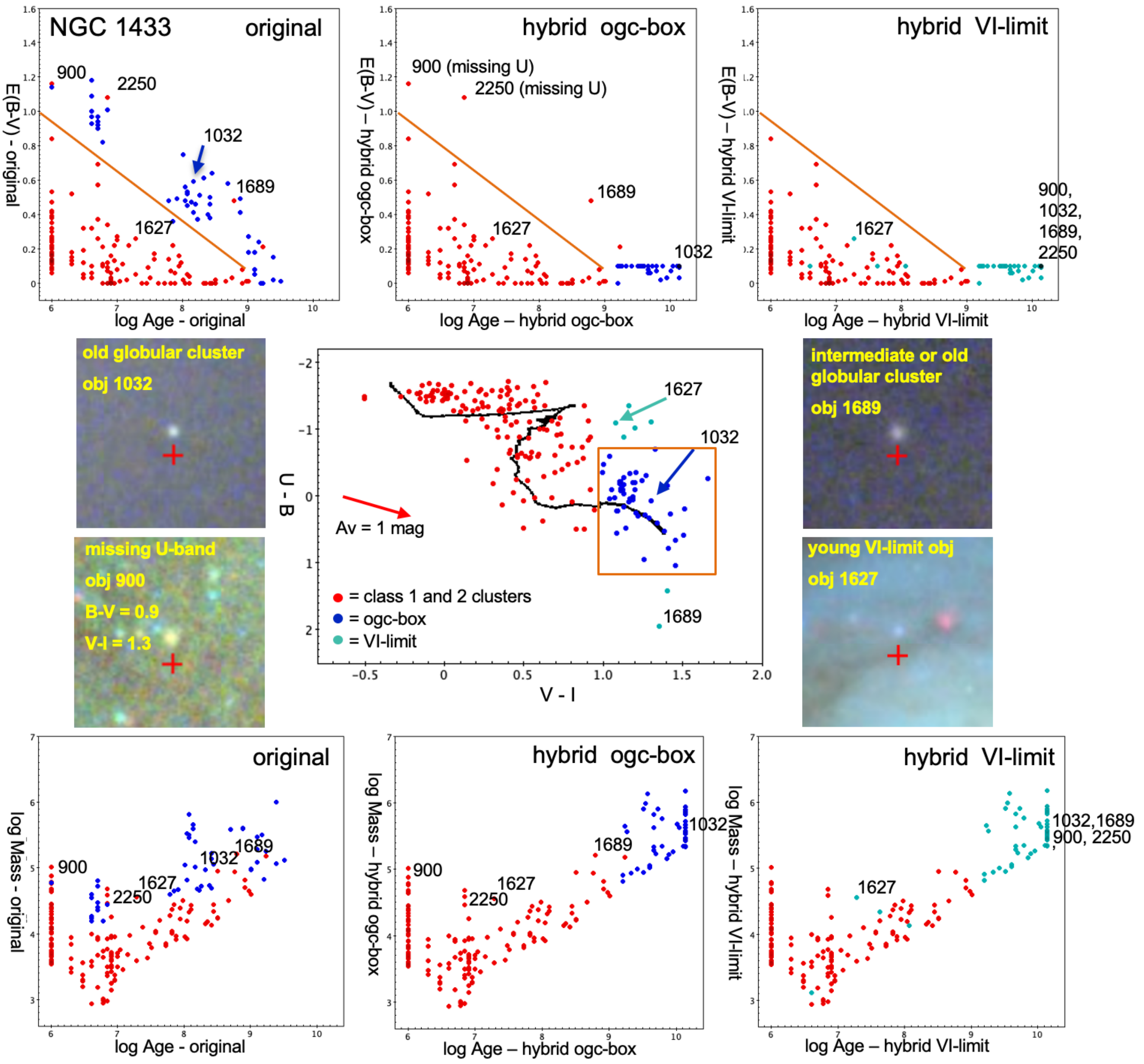}
\end{center}
\caption{Figure illustrating most of the key points from the paper. The left panels show the original results from version 1 of the PHANGST-HST pipeline \citep{turner21}, the middle panels show results based on the hybrid OGC-box method, and the right panels show the results using the hybrid VI-limit method.  
Blue points are found within the Old Globular Cluster (OGC) box shown in the $U-B$ vs. $V-I$ diagram in the central plot. Blue and cyan points are for the OGC-box and VI-limit solutions respectively.  Log Age vs. $E(B-V)$ (i.e. reddening) plots are shown in the top panels for the original and the two hybrid solutions. Corresponding log Age vs. log Mass plots are shown in the bottom row. Four snapshots of illustrative clusters are shown with their object numbers. These numbers are included throughout the diagram to show how their positions change in  the three different solutions.  }
\label{fig:7plot-n1433}
\end{figure*}

\medskip

 %For the first three galaxies (
For  NGC 1433, NGC 3351, and NGC 628 we use three additional steps to fix most of the incorrect globular cluster ages:

% \begin{enumerate}
%     \item 
  1).  Identify clusters which 
%    potentially have bad ages 
    are likely to be old globular clusters due to their position in the $U-B$ vs. $V-I$ color-color diagram (i.e., the OGC-box solution, where 
    $-0.7 < U-B < 1.2$ and $0.95 < V-I < 1.7$), 
    or their $V - I$ value (i.e., the VI-limit solution, where $V-I > 0.95$).
    %%% - this works better when the U-band measurement is uncertain or missing).
%    from the PHANGS pipeline using simple color selections
    
%     \item 
    2). Fix ages for the identified clusters using a low-metallicity (1/50th solar) model and low maximum allowed $E(B-V) < 0.1$~mag, appropriate for most old globular clusters. 
    
%     \item 
3). Identify potential young interlopers (i.e., young clusters with enough dust to give them colors similar to old globular clusters) using either H$\alpha$ flux (i.e., with  H$\alpha > 3.0 \times 10^{-16}$ erg/s/cm$^2$/pixel, where pixel refers to MUSE pixels), CO intensities, or the presence of dust based on the HST image (see Section \ref{sec:halpha_co_dust}). We use ages from the original solar metallicity model for these young interlopers. 
%   \end{enumerate}
  
  \medskip
    For the very dusty NGC 1365 we use a slightly different method:

% \begin{enumerate}
    
%    \item 
1). Use a smaller OGC-box to identify old globular cluster candidates (i.e., the orange box in the bottom right panel of Figure \ref{fig:cc_4plot};  $-0.6 < U-B < 0.5$ and $0.95 < V-I < 1.5$). We use the same methods described in steps 2 and 3 above to infer appropriate ages for these objects.  
    
%    \item 
2). Identify the high reddening objects (i.e., V-I $>$ 1.5), which are nearly all young, dusty clusters, and use a solar metallicity model, {\it but require higher values of $E(B-V)$} (1.5 $<$ $E(B-V)$ $ 2.5$~mag). This allows the reddest, dustiest clusters to reach the young ages that are correct for them. 
    
%    \item 
3). Identify "old interlopers" present in the sample of high reddening objects using H$\alpha$ flux (i.e., H$\alpha < 3.0 \times 10^{-16}$ erg/s/cm$^2$/pixel - note the inequality is in the opposite sense compared to young interlopers), CO intensities, or the presence of dust. We use the  original solar metallicity model solution for these objects.

 \medskip
 
 In principle, we expect H$\alpha$ to provide the best means of identifying interlopers, since it is generated by photo-ionization driven by massive young stars in the clusters themselves, while CO and dust
 are found in region that are statistically more likely to have 
 %higher likelihood of containing 
 young stars and clusters. We will revisit this question in Section \ref{sec:dust_halpha}.

More details used to illustrate these steps are included in the next subsection. 
We will refer to the results from these procedures as 
%% any set of results that combines corrected solutions for some clusters with the default PHANGS-HST  pipeline results for others as 
'hybrid' solutions.
It is important to note that the results presented in this paper are specific to the PHANGS-HST pipeline, including filters, assumed SSP model, etc, and that other works may have different numbers of objects with bad ages and different levels of improvement.

%\subsection{Hybrid solutions for NGC~1433 - improvement by 19\%}
\subsection{Solutions for NGC~1433 - improvement by $\approx$18\% in the overall success fraction}
\label{sec:hybrid-1433}

Figure \ref{fig:7plot-n1433}  shows an example of the primary  diagram that will be used in  this section of the paper to illustrate the success of the hybrid solutions. 
We show two different hybrid solutions: one based on the OGC-box, and one on the VI-limit solution. The VI-limit solution turns out to be slightly superior  for all four galaxies. The OGC-box approach is retained in the paper to help demonstrate a number of important points. 

The top left panel shows the log Age vs. $E(B-V)$ diagrams for the original pipeline solution; the top-middle panel shows the hybrid OGC-box solution; and the top-right panel shows the hybrid-VI-limit solution. The central plot shows the $U-B$ vs $V-I$ color-color diagram, with the blue objects in the OGC-box highlighted throughout the figure. The bottom panels show the log Age vs. log Mass diagrams. Snapshots of various objects are also shown and labeled  throughout  the diagram to illustrate various points.
%% and OGC-box clusters in blue (see middle panel) throughout Figure~\ref{fig:7plot-n1433}.  

We first note that clusters with colors within  the OGC-box clearly dominate the region above the diagonal line in the original log Age vs. $E(B-V)$ diagram, with one clump at very young 
% (log~$(\tau/\mbox{yr}) \lea7$) ages, 
 ages (log Age $\lea$ 7),
and a second near log Age $\approx$ 8. It turns out nearly every single one of these clusters was identified as having a bad pipeline age through the independent visual examination described in Section \ref{sec:approach} and shown in Figure \ref{fig:n1433_cc_bad12}. 

We next examine the top-middle panel that shows  results using the  OGC-box solution. Here, the OGC-box clusters are no longer above the diagonal line, and all but three objects  have moved to ages older than log~Age = 9 where they belong, with the largest concentration having log~Age = 10.1. The number of old globular clusters (i.e., with log Age greater than 9.5) has gone from 1 to 43; a much more realistic result (e.g., see \citealp{lomel22}).

We find that the limiting issue for the hybrid OGC-box approach is the lack of reliable U-band photometry, which appears to be the issue for the three objects (\#~900, \#~1689, and \#~2250) which remain above the diagonal line when using this approach.
%, namely objects 900 (negative flux in F336W), 2250, (negative flux in F336W) and possibly 1689 (F336W = 25.80 +/- 1.27 mag object). One of the remaining objects is cluster 1689, as shown in the small image on the right. 
The snapshot of \#~1689 shows this it is an old globular cluster, but with colors outside the OGC-box (possibly due to large U-band uncertainties), so its age has not been corrected.
This lead us to identify old globular clusters using a simpler $V-I$ $>$ 0.95 cut, as shown by the cyan points and the upper-right panel. All three of the clusters remaining above the diagonal from the OGC-box solution now move to older ages, and we reach a 40$/40=100$ \% success rate for the objects above the diagonal!  Hence the VI-limit approach to identify potential bad ages works extremely well for  NGC~1433. This is because there is insufficient dust to redden young objects enough to reach the OGC-box.

There are 191 total Class $1 + 2$ clusters in Figure \ref{fig:7plot-n1433}.  If we assume that all objects below the diagonal have correct ages,  the hybrid OGC-box solution goes from an  overall correct age fraction of 
79 \% = (191 - 40) / 191 = 151/191, to 98 \% = (191 - 3) / 191 = 188/191.
%% an improvement of 19 \%. 
For the hybrid $V-I$-limit solution the result is 191/191 = 100\%;  a 21 \% improvement from the original solution.  

We have checked our assumption that all clusters below the diagonal  have correct ages by manually examining
all the objects in NGC 1433. Based primarily on a visual examination of the HST image for the  presence or absence of H$\alpha$ emission, we estimate that three objects appear to have ages that are overestimated while three others are underestimated, hence the correct percentage below the diagonal is (141 - 6) / 140 = 96 \% for NGC 1433.
Taking this into account would therefore only reduce the overall success fraction by about 3 \%, as summarized in Table \ref{tab:table1}.

A comparison of the bottom-left  panel in Figure \ref{fig:7plot-n1433} with the two hybrid solutions to the right shows that 
%that A comparison of the original mass function (bottom left panel) and the hybrid ogc-box solution mass function (bottom middle panel) shows that 
roughly 20  clusters have changed age estimates from log Age $\approx 8$ 
%% (log Age = 6.8) 
to older than log Age  $\approx 9$, and most of these are at the high mass end. Therefore the improved age solutions from the OGC-box and VI-limit approaches will likely affect the overall age and mass distributions at a significant level. We will examine the impact of age corrections on the NGC~628 cluster population in  Section \ref{sec:age_and_mass}.

\begin{figure*}
\begin{center}
\includegraphics[width =7in , angle= 0]{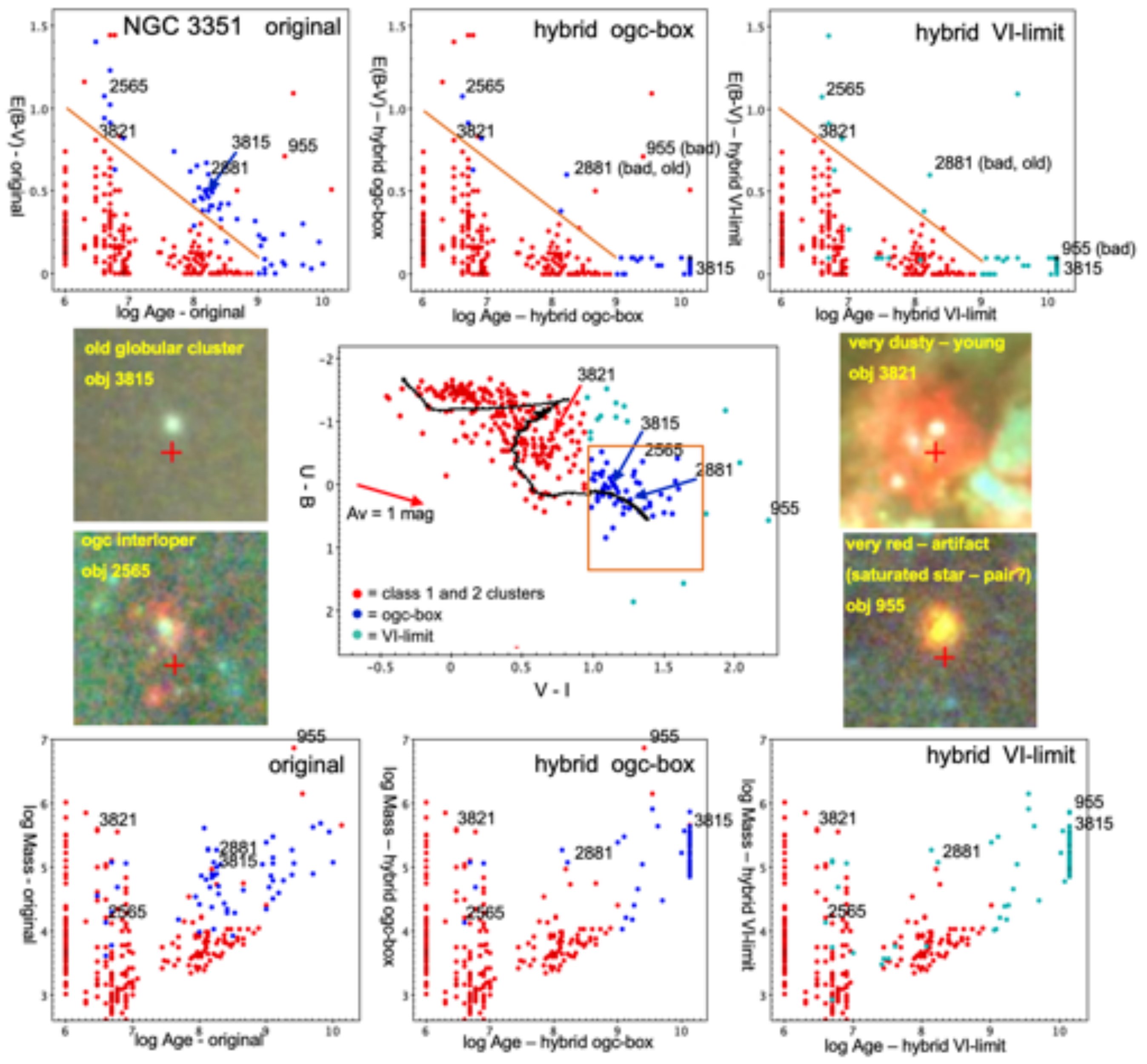}
\end{center}
\caption{Same as Figure \ref{fig:7plot-n1433} for NGC 3351. }
\label{fig:7plot-n3351}
\end{figure*}

\subsection{Solutions for NGC 3351 - improvement by $\approx$9 \% in the overall success fraction}
\label{sec:hybrid-3351}

Figure \ref{fig:7plot-n3351} shows
results for NGC~3351, which has the second lowest SFR in our sample but a fair amount of dust, especially in the inner starburst ring. 
We find 41 of 317 clusters have pipeline age-dating results that put them above the diagonal in the top-left panel. These clusters have questionable ages, as discussed earlier.  The log~Age vs. $E(B-V)$ plot when ages of clusters in the OGC-Box (blue points) have been corrected is shown in the top-middle panel. While most of the blue points have moved from above the diagonal to the bottom right, appropriate for old globular clusters (e.g. object \#~3815), 12  points remain above the diagonal. 
%above the diagonal for NGC 3351. 
These are evaluated below, and the numbers are summarized in Table \ref{tab:table1}.

\begin{figure*}
\begin{center}
\includegraphics[width =7in , angle= 0]{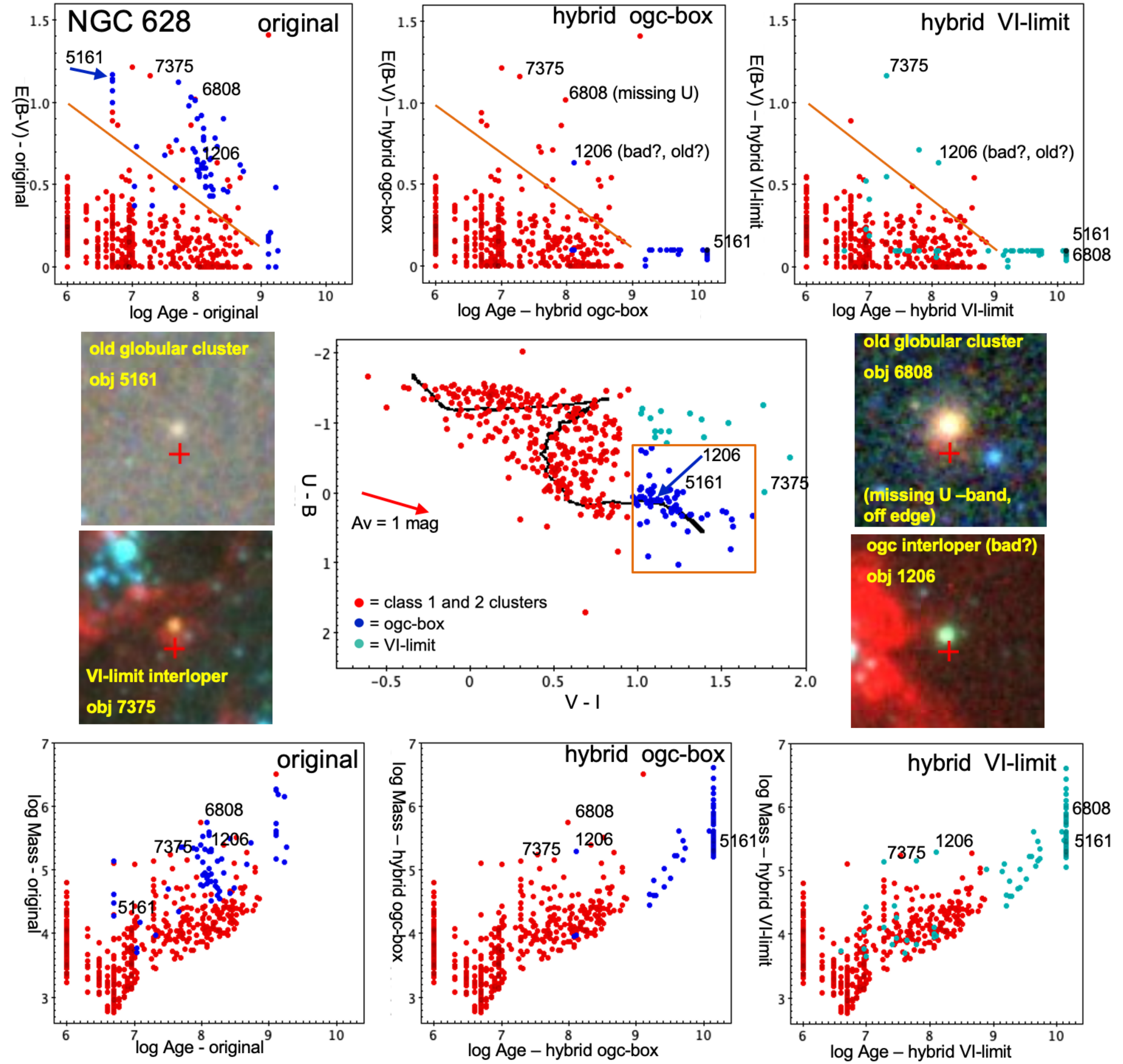}
\end{center}
\caption{
Same as Figure \ref{fig:7plot-n1433} for NGC 628. 
}
\label{fig:7plot-n628}
\end{figure*}

Four of the six red points (i.e., with colors that are not in the OGC-box) above the diagonal in the top middle panel have very faint or missing U-band measurements, so are not  identified as potentially having bad ages using the OGC-box selection because of the lack of a $U - B$ value.
%are similar to the objects remaining in the hybrid-OGC solution for NGC 1433, namely clusters with very faint (i.e., uncertain) or missing U-band measurements. These result in bad ages for the hybrid-OGC solution since they are not in the OGC-box because of the lack of reliable U-band measurement, 
The ages for these clusters do get fixed (changed from young ages to old), however, if we select objects using the VI-limit method (top -right panel). This is the primary reason the hybrid VI-limit solution looks better, with only seven points remaining above the diagonal.  

%Unlike NGC 1433, with very little dust, NGC 3351 has a fair amount of dust, especially in the starburst ring structure in the inner region. This results in 
On the other hand, there are five young "interlopers" in the OGC-box and VI-limit solutions, since they have values of H$\alpha$ $>$ $3.0 \times 10^{-16}$ erg/s/cm$^2$/pixel (e.g., see snapshot of \#~2565).
 These are the five blue or cyan points that remain above the diagonal in the top middle and top-right panels.
 %plot which also remain in the top right (VI-limit) panel. 
 One of these objects (\# 2881) turns out to be a "false interloper": the lower resolution MUSE observations indicate this cluster has H$\alpha$ emission (picked up from nearby regions), while the higher resolution HST H$\alpha$ observations shows there is no associated line emission at the location of the cluster itself.
 %due to the poor resolution of the MUSE H$\alpha$ observations, which picks up emission from nearby regions but does not have any H$\alpha$ flux associated with itself, as shown if one looks at the HST image. 
 This "resolution problem" is one of the primary failure modes of our current approach, and happens in approximately 20~\% of the cases where the H$\alpha$ flux indicates the presence of a young interloper. Future, high resolution H$\alpha$ observations from HST for the entire PHANGS galaxy sample, would solve this particular problem, and significantly improve the age-dating. H$\alpha$ observations using HST are now planned for 19 of the 38 galaxies in cycle 30 (i.e., the MUSE galaxies in the PHANGS-HST sample: proposal 17126 - PI = Chandar).

The remaining objects above the diagonal in the hybrid OGC-box solution are more of a mixed-bag, as summarized in the notes to Table \ref{tab:table1}. They include a background galaxy, another object with the resolution problem discussed for object \# 2881, and an object with $V - I$ = 0.94 (i.e., just missing the color cutoff for the OGC-box).

Turning our attention to the log Age vs log Mass diagrams in Figure \ref{fig:7plot-n3351}, we can see that many clusters now have shifted from estimated ages between 8 $<$ log Age $<$ 9 to older ages with log Age $>$ 9.  The apparent deficit of clusters between 10 and 50 Myr is
a well-known age-dating bias, which results when the model predictions reverse direction and loop back on themselves, giving similar predicted colors over a broad age range (e.g., \citealp{chandar10b}).  The resulting `gap' is observed at some level in all four galaxies.  This particular artifact only shifts age estimates by up to $\pm 0.3$ in log age,  significantly less than the factor of ten we define as "catastrophic" errors in our accounting.

The total number of clusters in Figure \ref{fig:7plot-n3351} is 317.  If we again assume that ages for all objects below the diagonal line are correct, and four of the objects above the diagonal have good ages (as indicated from the VI-limit solution)   the success fraction  goes from being %overall correct age ratio of 
88~\% correct
% =  (317 - 41 + 4) / 317 = 280 / 317
for the original pipeline solution,
to 97 \%  
% = (309 / 317) 
for the hybrid OGC-box solution, and 99 \% 
% = (314 / 317)  
for the hybrid VI-limit solution (this solution drops to 97\% if we account for a handful of incorrect age estimates below the diagonal line based on our review of NGC~1433). Hence we have improved the overall age dating by $\approx$~9~\%.   
%The bottom plot in Figure \ref{fig:7plot-n1433} shows the log Age vs log Mass diagram for the standard (solar) solution on the left and the hybrid solution on the right. 
Similarly the number of old globular clusters 
(i.e., with log Age $>$ 9.5) increases from just 7 to 44 for the hybrid OGC-box solution, and to 50 for the hybrid VI-limit.

\subsection{Solutions for NGC~628 - improvement by $\approx$9 \% in the overall success fraction}
\label{sec:hybrid-628}

Figure \ref{fig:7plot-n628} shows the 
results for NGC~628, the galaxy with the second highest SFR in our sample. We find 70 out of 489 data points above the diagonal in the standard model in the top left panel, which represent questionable  age estimates. 
As we found for NGC~1433 and NGC~3351, many of the objects left above the diagonal line after identifying and correcting ages in the OGC-box are those  missing U-band measurements (11 of 23), although for NGC~628 most of these are because they are outside the field of view.
This  is why the VI-limit model is more successful at moving points above the diagonal to older ages than the OGC-box method for  NGC~628.

There are only three remaining young interlopers in NGC~628 for the VI-limit model based on H$\alpha$ emission values greater than $3.0 \times 10^{-16}$ erg/s/cm$^2$/pixel (i.e., the cyan points above the diagonal in the upper right panel).
One of these (object \# 1206)
is questionable, since it is near the  outskirts of a
% Object \# 7375 (see snapshot in Figure \ref{fig:7plot-n628}), has V-I = 1.75, which means it would be selected as a globular cluster by the VI-limit method but not by the OGC-box method. 
% The second interloper (OGC) is source \# 1206. This one is harder to age-date because it is near, 
nearby young association (see snapshot in Figure \ref{fig:7plot-n628}), but may not be associated with it.

\begin{figure*}
\begin{center}
\includegraphics[width =7in , angle= 0]{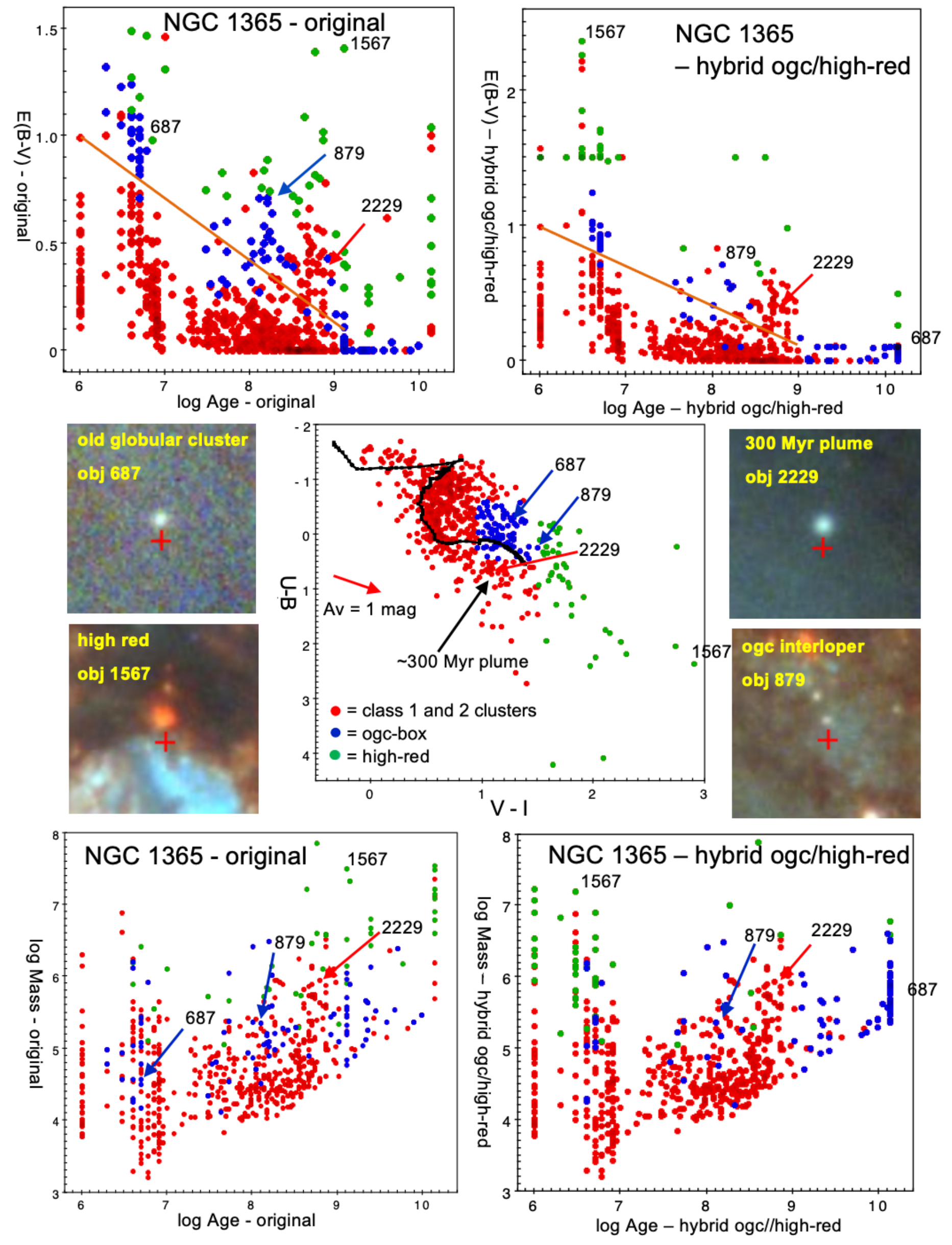}

\end{center}
\caption{Similar to Figure \ref{fig:7plot-n1433} for NGC 1365. The primary difference is that the hybrid OGC/high-red solution is shown rather than the OGC-box and VI-limit solutions. In addition, the OGC-box is smaller, and green points are used to show  the high-red clusters. See text for details. }
\label{fig:7plot-n1365}
\end{figure*}

The next set of objects ($N = 8$) we consider are barely above the diagonal line in the upper left panel and have V-I colors just blueward of the $V-I$ $>$ 0.95~mag cutoff, ranging from 0.75 to 0.93~mag. Hence these objects are not selected by either the OGC-box or VI-limit criteria for age-correction. 
Seven of the eight are assigned log Age $\approx 8.5$, which appears to be correct based on the absence of  H$\alpha$ emission in these objects.  Hence the placement of the cutoff at $V-I$ $>$ 0.95~mag appears to be accurate these objects. 
Table \ref{tab:table1} summarizes the various objects above the diagonal, and whether they appear to be good or bad age estimates.

If we assume that all ages below the diagonal are correct, plus nine additional objects above the diagonal are correct as discussed above,
%(e.g., as indicated from the VI-limit solution)   
the success fraction  improves from %overall correct age ratio of 
88~\% 
% =  (489 - 70 + 9) / 489 = 428 / 489
for the original pipeline solution,
to 98~\%  
% = (479 / 489) f
for the hybrid OGC-box solution, and 99~\% 
% =  (485 / 489) 
for the hybrid VI-limit solution (97~\% if we account for a few incorrect age estimates below the diagonal line based on our NGC 1433 visual review). Hence we have improved the success fraction by $\approx$ 9 \%.
% The bottom panels in Figure \ref{fig:4plot-n1433} 
%shows the log Age vs log Mass diagram for the standard (solar) solution on the left and the hybrid solution on the right. The 
The number of old globular clusters (i.e., with log Age greater than 9.5) increase from none in the original pipeline solution to 53 (65) for the hybrid OGC-box (hybrid VI-limit) methods, again yielding much more realistic figures.

\subsection{Solution for NGC~1365 - improvement by $\approx$13 \% in the overall success fraction}
\label{sec:hybrid-1365}

Figure \ref{fig:7plot-n1365} shows the 
results for NGC 1365, which has the highest SFR in our sample and in the entire PHANGS-HST sample \citep{lee22}.  We find 128/635 = 20 \% of the data points are above the diagonal in the top left panel. Unlike the other galaxies, most of these clusters belong here, since they are young and dusty.
% many more than for the other three galaxies we have observed. This is due to the presence of extensive dust in the galaxy, and makes the age dating of this galaxy much more difficult.
Because of this,
%dust and other attributes that may or may not be  unique to NGC 1365 (i.e., the "300 Myr plume" which trails down the reddening vector from about the 300 Myr point on the SED), 
we use a somewhat different approach for correcting bad cluster ages in NGC~1365, as described in Section \ref{sec:approach}. 

The first difference is how we select potential old globular clusters. We use a smaller OGC box, with $-0.6 < U-B < 0.5$   and $0.95 < V-I = < 1.5$  to avoid the `300 Myr plume' just below the OGC-box in Figure \ref{fig:7plot-n1365}. This is discussed in more detail in Section \ref{sec:300myr}. We then identify clusters redder than $V-I = 1.5$ on the right side since these very red clusters appear to nearly all be young, unlike the other galaxies. Their youth is indicated by strong CO emission (black points) for nearly all clusters with $V-I$  measurements $>$ 1.5~mag
in Figure \ref{fig:cc_4plot}. We  call these the "high-red" objects in what follows. 
%, where nearly all clusters with V-I  measurements between 1.5 and 1.7~mag in the normal OGC box are black, indicating strong CO emission. 
The second change allows the SED-fitting routine to reach the correct young age solution for the highly reddened, young clusters  by restricting $E(B-V)$ values to be in the range 1.5 - 2.5~mag
%accounts for the  large number of highly reddened, young clusters by adopting $E(B-V)$ values between 1.5 and 2.5~mag, to allow the SED-fitting routine to reach the correct young age solution
% with 11 clusters with V-I $>$ 2 or U-B $>$ 2. 
(i.e., addressing Problem \# 3 in Section \ref{sec:common_problem_3}, and using the procedure described in Section \ref{sec:procedure}). 
% The large population of highly reddened, young clusters is the reason there are so many apparent old globular clusters with high reddening in the original pipeline solution shown in the upper left panel in Figure \ref{fig:4plot-n1365}.
%, i.e. the green points defined by V-I $>$ 1.5. 
A visual examination of these high-red objects confirms that nearly all are very young, with strong H$\alpha$, strong CO, and large amounts of dust around them. 

Cluster \# 1567  (see Figure \ref{fig:7plot-n1365}) illustrates the problem. It is in the very dusty central region of NGC 1365, has strong H$\alpha$ and CO emission, and lots of young blue $\approx$few~Myr objects nearby. 
The original pipeline SED solution assigned it a log Age = 9.1 and a reddening $E(B-V)$$=1.47$~mag (near the adopted maximum of 1.5~mag). 
%However, it is found in the central region of NGC 1365 in a very dusty region, with very high values of H$\alpha$ and CO, and lots of young blue objects with ages of only a few Myr nearby. 
For the solution to reach its likely age of $\approx 5$ Myr, an $E(B-V)$ around 2.5~mag is required. So for clusters with very red V-I colors in NGC~1365 (i.e., the green points in Figure \ref{fig:7plot-n1365}), we adopt SED-fitting solutions with solar metallicity and values of $E(B-V)$ between 1.5 and 2.5~mag. This fitting finds a solution of log Age = 6.48 and $E(B-V)$ = 2.36, for object \# 1567, a much better estimate for this cluster.  Most of the other highly reddened objects that were originally given old ages are now correctly fit with ages $\approx 5$~Myr plus high reddening.
%this allows the object to be assigned a  value of log Age = 6.48 and $E(B-V)$ = 2.36, which appears to be a much better estimate for the object. 
%The same happens for most of the other objects that were original given ages of old globular clusters; i.e., most are assigned ages around 5 Myr.

Because most of the  objects in the high-reddening subsample  are young rather than old,  we switch the direction of the inequality used to identify "young interlopers" in the OGC-box and now identify "old interlopers" in the high-reddening subsample instead. These are defined as objects with weak H$\alpha$ flux (H$\alpha <  3.0 \times 10^{-16}$ erg/s/cm$^2$/pixel - i.e., old globular clusters). 
In these cases we use the original SED solution for the age, $E(B-V)$, and mass of the cluster.  This leads to only 8 of the 43 high-red clusters being assigned ages older than log Age = 7.5; the other 35 are all assigned ages younger than 10 Myr (upper-right panel in Figure~\ref{fig:7plot-n1365}), which is supported by  visual examination.

NGC 1365 has essentially the same number of clusters above the diagonal after age corrections are made (i.e., N = 120), as before (i.e., N = 128), unlike the first three galaxies where most of the objects above the diagonal are moved to older ages in the hybrid solutions.  While roughly 30 old globular clusters change from intermediate pipeline ages to old ages (blue points), an approximately equal number of clusters which were originally given incorrect old and intermediate ages by the PHANGS pipeline have hybrid solutions which move them in the opposite direction, to very young ages (green points).
The bottom plots in Figure \ref{fig:7plot-n1365} show the log Age vs. log Mass diagram for the standard (solar) solution on the left and the hybrid solution on the right. The number of old globular clusters (i.e., with log Age greater than 9.5) in the hybrid solution
%% has gone from 22 - 8 [high red] = 14  for the original solution to 
is 55, similar to the other three galaxies in this pilot study.    

A possible contaminant in our sample of very red clusters might be background galaxies. Since our candidate clusters have been selected based on human classifications \citep{whitmore21} we expect this to be a  rare occurrence. In addition, a secondary visual check of all the cluster candidates in the inner region of NGC 1365, where the reddest objects are found, resulted in only one object that appeared to be a possible background galaxy.

The total number of clusters in Figure \ref{fig:7plot-n1365} is 635.  It is more difficult to estimate the improvement in the success fraction for NGC 1365 since objects are moving both into and out of the region above the diagonal line. However, a rough estimate can be made based on the census of good and bad ages in Table \ref{tab:table1}. This leads to estimates of 
84~\% 
% = 635 - 128 + 24) / 635 = 531 / 635 
for the original pipeline solution and 97~\% 
% =  (635 - 120 + 100 / 635 = 615 / 635 f
for the hybrid ogc/high-reddening estimate, an improvement of 13~\%.

In addition, our estimate of 96~\% for the success rate below the diagonal based on a visual check in NGC~1433 is clearly too high for NGC~1365 due to the high level of chaotic dust in the inner region of this galaxy. 
A spot check of the success fraction below the diagonal line indicates that a value of 90~\% is more appropriate. In this case, our overall success fraction would fall from  97~\% to about 92~\%, as reported in Table \ref{tab:table1}. 

The recent  JWST 
observations of NGC 1365 
(\citealp{whitmore22}, 
% THIS DOESNT SEEM TO WORKK YET - DEC 9/2022 >>> \citep{WHITMORE_PHANGSJWST}
\citealp{lee22b}) provide an additional check on our age estimates. One result from this work is the finding that 21 micron flux is an excellent predictor of whether a cluster is old (faint or missing at 21 micron) or young (bright at 21 micron). A check of all young class 1 and 2 clusters from the human classified compact cluster catalog \citep{whitmore21} with hybrid OGC/high-red age estimates log Age $<$ 7 and log Mass $>$ 6, finds that 21/24 = 88~\% clusters have a value of 21 micron brightness $<$ 23 mag (abmag, 0.15 arcsec radius aperture, see \citealp{whitmore22} for details), as expected. Similarly, 17/20 = 85~\% of the old log Age $>$ 9.5 and log Mass $>$ 6 clusters have a a values of 21 micron brightness  $>$ 23 mag, as expected. Combining these two  estimates yields 38/44 = 86~$\pm$~6 \%, in reasonable agreement with the estimate of 92~\% accuracy in Table \ref{tab:table1}.

\begin{figure*}
\begin{center}
\includegraphics[width =7in , angle= 0]{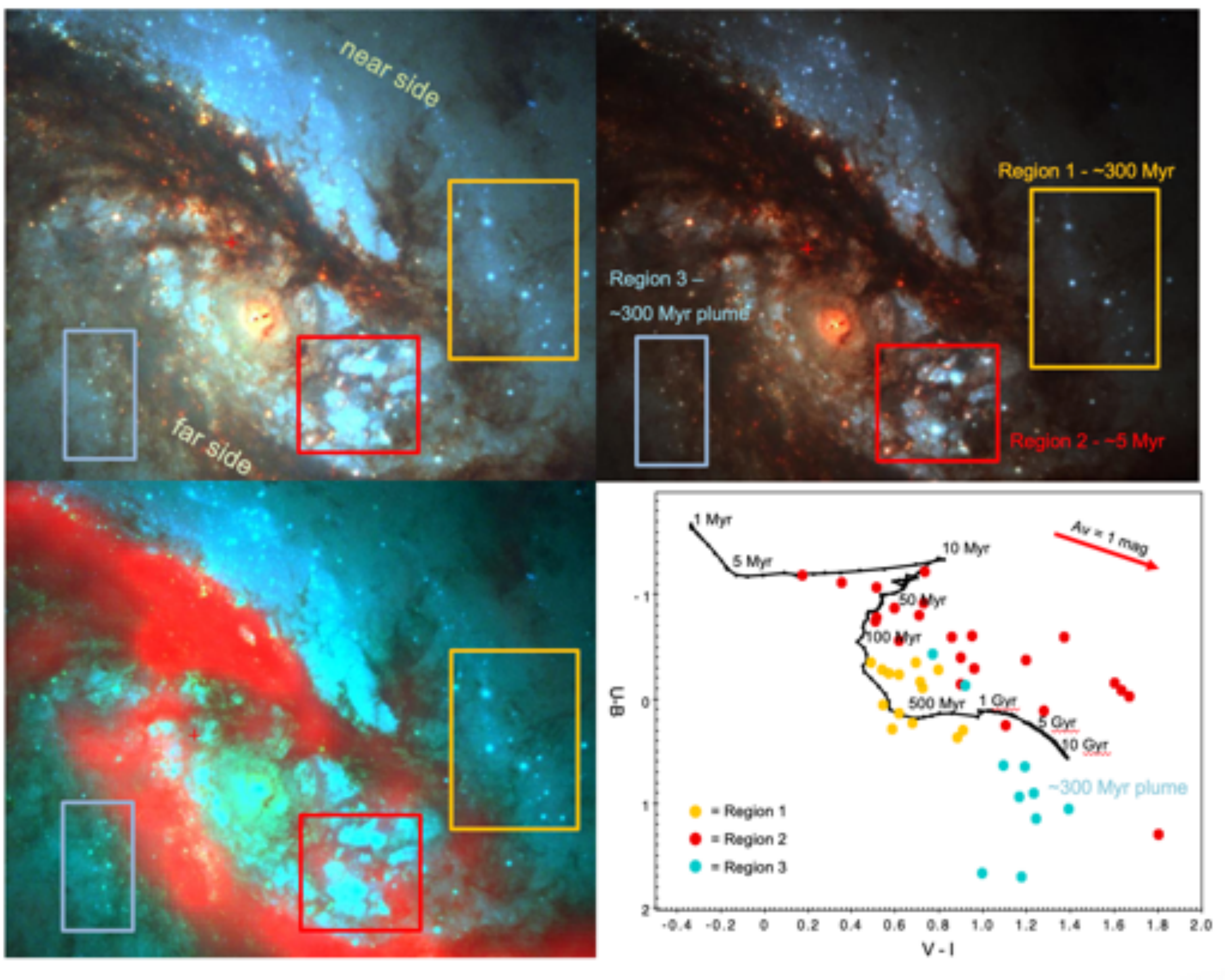}

\end{center}
\caption{Three regions in the inner part of NGC 1365. The upper two plots are HST B-V-I images  with different contrast levels; the bottom left images is a B-V-CO image. Regions 1 (yellow) and 3 (blue) are examples of 300 Myr clusters around the edges of the dust lane, as discussed in the text. Region 2 (red) shows a region of active star formation with a mean age $\approx5$ Myr. The color-color diagram in the bottom right shows the associated, color-coded locations for the clusters in the three regions. Note the location of the light blue points in what is called the 300 Myr plume in Figure \ref{fig:7plot-n1365}. This is caused by larger amounts of dust in front of these  objects when compared to Region 1 where there is little dust.    }
\label{fig:300myr_plume}
\end{figure*}

\subsection{The 300 Myr `plume' in NGC 1365}
%% of Reddened, Young Clusters}
\label{sec:300myr}

As briefly mentioned above, and shown in Figures \ref{fig:cc_4plot}, \ref{fig:7plot-n1365}, and \ref{fig:300myr_plume}, another somewhat unique characteristic of NGC~1365
%(although it may be common in other barred galaxies as discussed below), 
is the large number of intermediate age clusters, % with ages  $\approx$300~Myr, 
% \mbox{few}\times100$~Myr,
many of which form a reddening `plume' trailing from a position on the SED track appropriate for 300 Myr clusters toward the bottom right of the $U-B$ vs. $V-I$ diagram.  A visual examination shows that most of these are located just outside strong dust lanes in the inner region of the galaxy, which explains their relatively high reddening values.
%(but not as high as young clusters closer to the , although not nearly as strong as objects closer to the nuclear starburst region.
There are approximately 60 of these $\approx$300~Myr clusters 
(log Age around 8.7 in the upper right panel in Figure \ref{fig:7plot-n1365}).

We believe this configuration of intermediate-age clusters is a feature of how barred spiral galaxies create new clusters (e.g., \citealp{sormani20}, \citealp{whitmore22}, \citealp{schinnerer22}).
%just outside the strong dust lanes with very active star formation is built into the way barred galaxies make new star clusters, as described by Emellesen et al. 2022. 
As the gas and dust move inward
%the central region of the galaxy it  
along the bar it triggers star formation, primarily in the outer edge of the dust lane. After the stars and star clusters form they fairly rapidly decouple from the dust and gas,
%they  slowly traverse slightly outside the gas and dust, 
since the stellar components have less dissipation than the gas. \citet{sormani20} estimates that it  only takes about 5 Myr to decouple.  While the gas spirals into the central region, creating the star-forming ring and inner disk, most of the stars and star clusters traverse slightly outside the region of gas and dust, populating what are called "overshoot" regions. The orbits of the star clusters formed in this way can take many shapes, as shown in Figure 8 of \citet{sormani20}. For example, a large fraction of the stars will follow X-shaped box orbits, precessing to occupy regions which are roughly 
45 degrees  from the plane of the inner disk after several orbits. 
% In addition, the apparent motion  of the star clusters will be slower when they are the outer parts of their orbits since they are moving tangentially to our line of sight, hence increasing the density still more. 
% This may be the reason we see the enhancements in the positions of intermediate-age clusters in regions 
This puts the intermediate-age clusters near, but not behind, the dust lanes for a large fraction of their orbit.

Figure \ref{fig:300myr_plume} shows two regions where intermediate-age clusters are enhanced, roughly 45 degrees from the plane of the inner star forming disk. The corresponding region in the color-color diagrams  are also shown. We find that all of the objects in Region 1 are consistent with being roughly 300 Myr intermediate-age clusters. In addition, the objects in Region 3 (the far side of the galaxy - see Figure \ref{fig:300myr_plume} and  \citealp{elmegreen09} ) appear to have more dust, as expected, hence explaining the higher reddening, as shown by their position in the 300 Myr plume in the color-color diagram. Other younger overshoot regions near the top of Figure \ref{fig:300myr_plume} are discussed in Whitmore et al. 2022 in press, and Schinnerer et al. 2022 in press.

A younger region in the inner star formation ring  (Region 2) is shown by the red outline and points in the color-color diagram. Most of these are  consistent with ages in the range 1 to 10 Myr, with values of reddening from zero to about A$_V$ = 4 mag.

\begin{figure*}
\begin{center}
\includegraphics[width =7in , angle= 0]{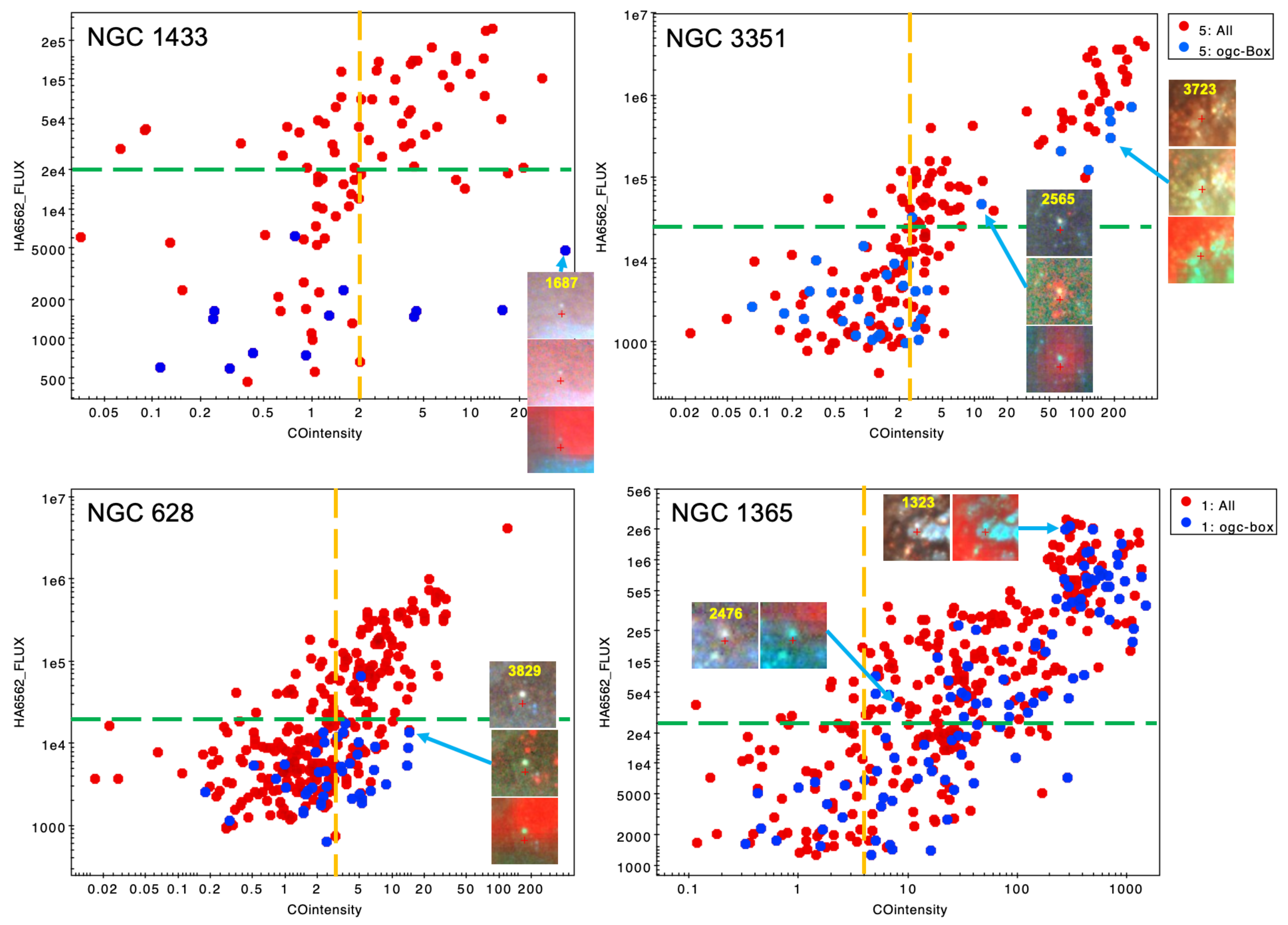}
\end{center}
\caption{
Measured   H$\alpha$ flux vs. CO intensity for the clusters in our four sample galaxies. In each panel, the green line shows where objects appear to be associated with  the faintest HII regions and the yellow lines shows the corresponding line for the CO flux of the same HII regions. These lines provide estimates of where the line should be set to determine whether an objects is an interloper, i.e., a young objects with enough dust that it has colors appropriate for an old globular cluster. Snapshots of various objects are included to illustrate various points in the text. The three snapshots are Hubble images using B-V-I  (top), B-V-H$\alpha$ (middle) and B-V-CO (bottom) images. The units for  H$\alpha$ flux are  10$^{-18}$ erg/s/cm$^2$/pixel and for CO intensity are K km s$^{-1}$.}

\label{fig:co_halpha_4plot}
\end{figure*}

\section{Using H$\alpha$, CO, or Dust to Identify and Fix Incorrect Ages of Interlopers}
%% of Reddened, Young Clusters}
\label{sec:halpha_co_dust}

We have shown throughout this work that young, reddened clusters and old globular clusters can have similar broad-band colors, and that additional information is needed to 
separate the two. 
% break this well-known degeneracy between age and reddening. 
In this section we explore how well the following star-forming tracers distinguish between the two cases: H$\alpha$ flux from MUSE, the CO intensity from ALMA, and the dust strength from HST images.

\label{sec:best-interloper}

\subsection{Comparison between H$\alpha$ and CO Flux}
\label{sec:halpha_co}

Here we assess whether H$\alpha$ flux or CO intensity is better at identifying young interlopers. 
% (i.e. reddened young clusters which have colors in the OGC-box or the VI-limit region of the color-color diagram similar ).  
We are fortunate to have high resolution, archival H$\alpha$ images from HST for three of our pilot galaxies (NGC 628, NGC 1433, and NGC 3351), which we use as a starting point.  Over half of the PHANGS-HST sample also have H$\alpha$ measurements from ground-based MUSE IFU observations, including all the galaxies studied here, so we will use the MUSE H$\alpha$ measurements as our primary tool for the broader sample, even though they are a factor of $\approx$10 lower resolution than the HST observations. 

Figure \ref{fig:co_halpha_4plot} shows the MUSE H$\alpha$ flux vs. ALMA CO intensity measured at the location of each cluster for all four galaxies (red points).  Note that a number of clusters are missing from this diagram because no CO flux is measured.  An examination shows that the missing sources are generally found between spiral arms or in the bulge, hence they are unlikely to be young interlopers in any case. 
 Clusters with broad-band HST colors that place them in the OGC-box (defined in Section~\ref{sec:approach})
%of a U-B vs. V-I color-color diagram 
are shown as the blue points. 
% The total number of clusters in the OGC-box are given in Table \ref{tab:table1}.

\subsubsection{NGC~1433}

We start by visually inspecting ten clusters in NGC~1433 with very faint H$\alpha$ emission (i.e. HII regions) in the HST images, to get a feel for the faintest detectable level of H$\alpha$ emission in the data.  Overall, we found a good correspondence between the presence of H$\alpha$ emission in the high resolution HST images and the lower resolution MUSE observations - when we see H$\alpha$ emission in one, we can also detect it in the other.  From this visual inspection, we set an approximate flux limit for the number of counts that separates clusters with and without H$\alpha$ emission (dashed green horizontal line in the upper left panel in Figure \ref{fig:co_halpha_4plot}); {\em clusters above this line are all expected to be young}.  In NGC~1433 this separation was originally set at $2.0 \times 10^{-16}$ erg/s/cm$^2$/pixel, as shown in Figure \ref{fig:co_halpha_4plot}. While there is some variation in the exact flux limit from galaxy to galaxy, 
we always find values close to $\approx3.0 \times 10^{-16}$ erg/s/cm$^2$/pixel, and hence adopt this value during final processing, as described in Section \ref{sec:approach}.

For NGC 1433, in the upper left panel of  Figure \ref{fig:co_halpha_4plot}, we find that no clusters shown as blue points from the OGC-box are H$\alpha$ emitters, because they are all below the green horizontal line. This is consistent with our discussion in Section \ref{sec:hybrid-1433}, where all NGC~1433 objects in the OGC-box appear to be old globular clusters: there are no reddened, young interlopers. 
%The green line in Figure \ref{fig:n1433_co_halpha} is in agreement with this prediction, and shows that 
A criteria of H$\alpha > 3.0 \times 10^{-16}$ erg/s/cm$^2$/pixel identifies young interlopers correctly 100 \% of the time in NGC~1433.

We make a similar estimate for CO intensity using the same ten clusters with weak H$\alpha$, and show the lowest value as the vertical yellow line.  Here, {\em clusters above a CO intensity value of 2 K~km~s$^{-1}$ are predicted to be young}. In this case, however, the CO measurements predict that four clusters in the OGC-box shown as blue points fall to the right of the line, i.e. they should be young interlopers, at least according to the CO intensity. Visual examination indicates that these are actually all likely to be old globular clusters, and the CO emission is not actually associated with the clusters.

%An example of what is happening is shown by the  
We illustrate this situation in the
snapshots of source \# 1687 in the upper left panel of Figure \ref{fig:co_halpha_4plot}.
%%% (see also Figure \ref{fig:4x2_image}).  
%% This turns out to be just below the CO/dust lane north of the  nuclear region of NGC 1433 
The top snapshot shows the B-V-I image. We find the object is isolated, and has a yellow color typical of old globular clusters. The middle diagram shows the H$\alpha$. We find weak H$\alpha$ emission (in red) spread evenly throughout the image, i.e., there is no evidence that the source has an HII region associated with it. 
The bottom snapshot  shows the CO intensity in red. We find that the object is on the edge of a strong CO region just as it intersects with the central CO ring of NGC~1433.  
Hence, at least in NGC~1433, our experiments indicate that H$\alpha$ is much better at identifying young, reddened interlopers than is CO. This makes sense, since H$\alpha$ emission is more closely associated with recently formed clusters than CO emission is (e.g., \citealp{chevance22}).

\subsubsection{NGC~3351}

\begin{figure*}
\begin{center}
\includegraphics[width =7in , angle= 0]{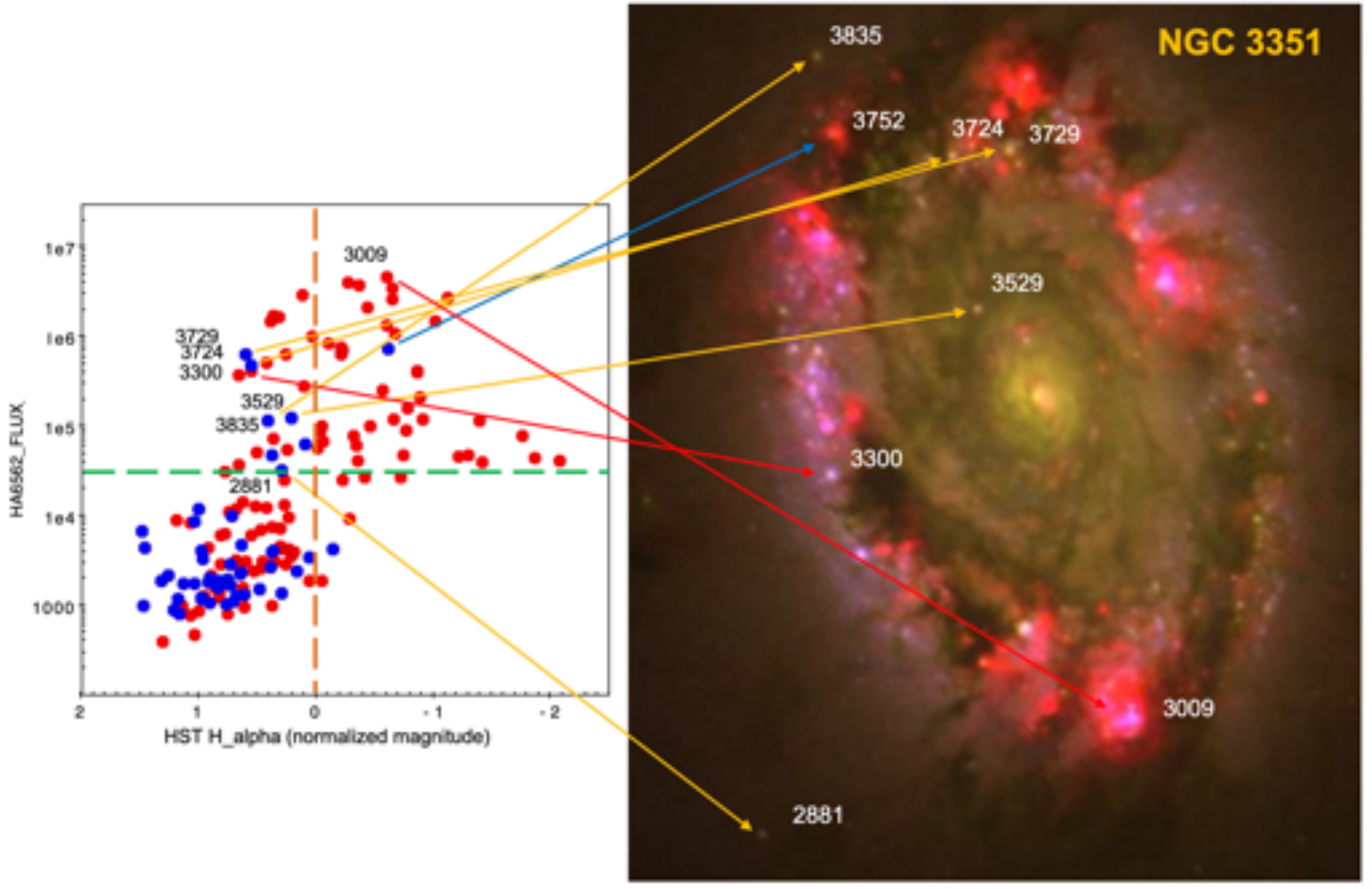}

\end{center}
\caption{Comparison of H$\alpha$ measurements from MUSE and HST, using the same conventions as Figure \ref{fig:co_halpha_4plot}. Several clusters are labeled and identified in a HST H$\alpha$/F814W/F438W image on the right. Note that while eight clusters would qualify as young interlopers based on MUSE observations (i.e., above the green dashed line), only two would qualify based on HST observations (i.e., to the right of the orange dashed line). See text for details. }
\label{fig:Halpha_2plot-n3351}
\end{figure*}

The upper right panel in Figure \ref{fig:co_halpha_4plot} includes the same diagram for clusters in NGC~3351. The distribution of cluster measurements looks quite different than in NGC~1433, being nearly bimodal rather than continuous.  This is due to clusters found in the inner star forming ring (see Figures \ref{fig:4x2_image} and \ref{fig:Halpha_2plot-n3351}) having higher CO and H$\alpha$ values 
%% (i.e., towards the upper right in the top-right panel of Figure~\ref{fig:co_halpha_4plot}) 
than nearly all the clusters in the outer part of the galaxy. 

The separation of H$\alpha$ (CO) emitting vs. non-emitting clusters shown as the green horizontal line (yellow vertical line) was determined using the procedure described above for NGC 1433, this time for NGC~3351.  In the lower `clump' of points, we find two sources in blue above the green line, which are predicted to be reddened young clusters (i.e., young interlopers) within the OGC-box.  We include snapshots for the object  \#~2565 (see also Figure \ref{fig:7plot-n3351}).
These show clear H$\alpha$ emission in the middle snapshot,  supporting the idea that this is a young, reddened cluster, or young interloper.  This object's location to the right of the vertical yellow line indicates that the CO measurement would also suggest it is a reddened, young cluster, as confirmed by the bottom snapshot for \#~2565. 

Next we examine object \#~3723 in Figure~\ref{fig:co_halpha_4plot}.  The three associated HST snapshots
%on the left (for the blue point nearest the intersection of the yellow and green lines) 
show the object is in a very dusty, H$\alpha$ and CO-rich area, and hence is a clear young interloper, as also appropriate for its location relative to the green and yellow lines. It is located in the strongly star-forming ring near the center of NGC 3351.

While these two objects would be correctly identified as young interlopers using both the green and yellow lines, we note the presence of five objects near the bottom of the panel for NGC~3351 that would be incorrectly identified as young interlopers by the CO intensities since they are to the right of the yellow line. 

\subsubsection{H${\alpha}$ measurements in NGC 3351: MUSE vs. HST}

We now assess how well the lower-resolution MUSE H$\alpha$ observations perform in identifying young interlopers compared with high resolution HST-H$\alpha$ images in the crowded star-forming ring of NGC~3351.
In Figure \ref{fig:Halpha_2plot-n3351} we compare the MUSE HA6562\_FLUX values from Figure \ref{fig:co_halpha_4plot} with HST F657N (H$_{\alpha}$) measurements made from 2 pixel aperture photometry. In order to (approximately) subtract out the continuum and identify sources with H$_{\alpha}$ line emission, we subtract the F814W HST magnitude, and  normalize the measurements so objects with no line emission 
%OGC-box objects in the outer regions of the galaxy nearly all 
have a positive F657N - F814W magnitude, as shown by the orange dashed line. 

The left panel of Figure \ref{fig:Halpha_2plot-n3351} shows that two clusters are identified as young interlopers based on HST H$_{\alpha}$ measurements, the blue points to the right of the vertical orange line.  One of these (\# 3752) clearly has strong H$_{\alpha}$ emission in the H$\alpha$-I-B  
% F657N/F814W/F438W 
image shown on the right, while the other is in the outer part of the galaxy and also shows clear but weaker line emission.
%does not show obvious line emission.  
These are the only two young interlopers within the OGC-box identified from higher resolution HST H$_{\alpha}$ imaging, and contrasts with the eight, mainly incorrect identifications, made from the MUSE HA6562\_FLUX (blue points above the green horizontal line).

We examine these sources in the high resolution HST image shown on the right in Figure \ref{fig:Halpha_2plot-n3351}.  Three of these clusters (2881, 3529, and 3835) have measured H$\alpha$ emission from MUSE but not from HST images.  All three appear to be old globular clusters and not young interlopers at higher resolution. 
%This is an improvement over using the MUSE H$_\alpha$ measurements.  
Two other clusters however, (3724 and 3729) have uncertain classifications, even at HST resolution. These do not have associated H$_\alpha$ emission, but are in crowded regions with other very young, H$_{\alpha}$ emitting sources.  The last 2 clusters are further out in NGC~3351 (not shown in Figure~\ref{fig:Halpha_2plot-n3351}), with H$_{\alpha}$ in the region but not directly associated with the source, and hence are unlikely to be young interlopers. For comparison, we point out two young, H$_{\alpha}$ emitting clusters which do not have colors in the OGC-box: cluster \# 3009 has an estimated age of 3~Myr and strong H$_\alpha$ emission in the HST image, and cluster \# 3300 is somewhat older at $\approx8$~Myr, and has  had time to clear out the H$_\alpha$ emission around it.

In summary, we find that higher resolution H$_{\alpha}$ observations from Hubble improve the age estimates for five of eight clusters in the OGC-box within NGC~3351 by showing that while there is H$_{\alpha}$  emission in the region, it is not directly associated with the clusters.  The age estimate for one cluster is identical between HST and MUSE, while for two others it remains ambiguous.  %hence there would be a fair but not dramatic improvement. 
NGC~1433 and NGC~628 have almost no clusters in the OGC-box with H$_\alpha$ emission detected by MUSE (none in NGC~1433 and only one in NGC~628), so the lower resolution H$_{\alpha}$ imaging does not have much impact in these galaxies.
%We note that there would be less improvement for NGC~1433 and NGC~628, since  there are fewer  objects in the  OGC-box with H$_{\alpha}$ to begin with (i.e., none in NGC~1433 and only one in NGC~628). 
There would likely be a major improvement in identifying young interlopers in NGC~1365 however.
%For NGC~1365, it is likely that there would be a major improvement, since there are a large number of objects with emission that could be better evaluated. 
We plan to use the PAH-emission from high resolution F335W JWST observations (\citealp{lee22b}, \citealp{whitmore22}) to improve cluster age-dating in NGC~1365.

\subsubsection{NGC~628}

The CO intensity vs. H$\alpha$ flux diagram for NGC~628 is shown in the lower left panel of Figure \ref{fig:co_halpha_4plot}.
It is fairly similar to the one for NGC 1433, 
but with even more clusters that would be incorrectly identified as young interlopers based on CO intensity.
%associated with high CO intensity that would be incorrectly identified as young interlopers. 
The snapshots of object \#~3829 show only a few small patches of H$\alpha$ which may or may not be associated with the cluster, and a position on the edge of strong CO flux, probably due to the resolution problem. 

One object of particular note is the object  in the far upper right of the panel.
%is the red point in the upper right, 
This source has the highest CO and H$\alpha$ emission by a large margin. This is the "Headlight Cloud", as studied in \cite{herrera20}.

\subsubsection{NGC~1365}

The lower right panel in Figure \ref{fig:co_halpha_4plot} shows the distribution of clusters for NGC~1365. As expected based on our earlier discussion, this population looks quite different from those in the other three galaxies, with a large number of young interlopers indicated by both H$\alpha$ and CO intensities.  

The snapshots only include B-V-I and CO for this galaxy since there is no HST H$\alpha$ image for NGC 1365. Object \#~3723 is a typical example of a young interloper with both strong H$\alpha$ and CO. 
Object \#~2476   shows a less obvious example 
% with extensive dust and CO. 
% Snapshots for object \#~2476 show an object  
with a relatively small amount of dust and likely foreground CO. Hence the designation of this object as an interloper is uncertain. This source is close to the crossing of the yellow and green lines used to define young interlopers, as might be expected for an object with an uncertain designation.

In the case of NGC~1365 there are roughly 20 objects near the bottom right of the panel that would be incorrectly identified as young interlopers by the CO intensities since they are to the right of the yellow line but have no clear H$\alpha$ emission  . Hence in all four of our target galaxies, using CO instead of H$\alpha$ to identify interlopers would result in more incorrect ages, ranging from 2~\% more for NGC~1433 and NGC~3351 to 4~\% more incorrect for NGC~628 and NGC~1365.

\begin{figure*}
\begin{center}
\includegraphics[width =7in , angle= 0]{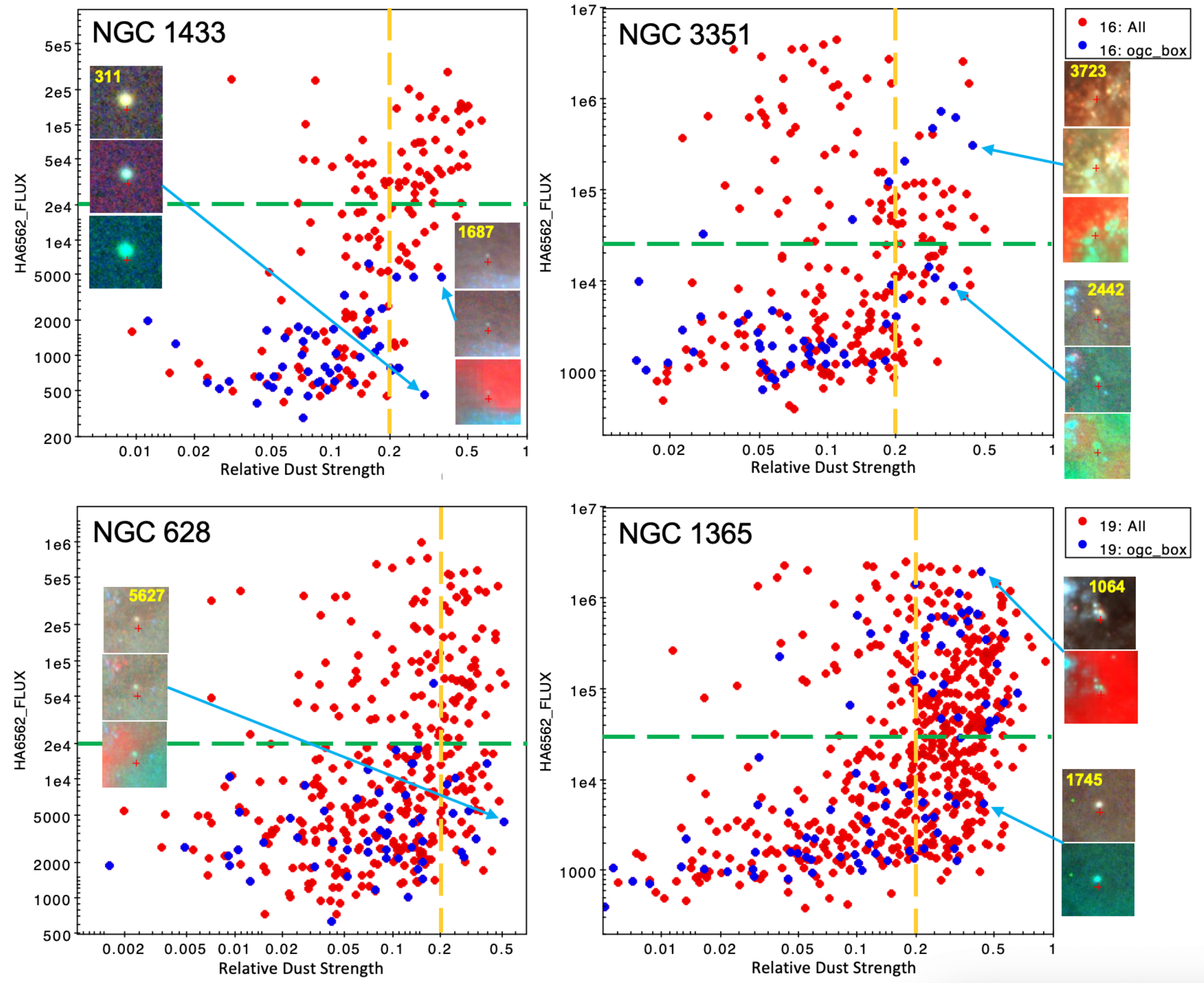}
\end{center}
\caption{Same as Figure \ref{fig:co_halpha_4plot} but for dust vs.H$\alpha$ for all four galaxies.  }
\label{fig:ha_co_4plot}
\end{figure*}

\subsection{Comparison between H$\alpha$ and Dust}
\label{sec:dust_halpha} 

In the previous section we found that H$\alpha$ emission is better at identifying reddened young interlopers that fall in the OGC-Box than CO emission at comparable resolution.  Here, we assess how well a measure of dust from high-resolution HST-based maps (\citealp{thilker23}
 - see Section \ref{sec:data})
compares with lower-resolution MUSE-based H$\alpha$ measurements at this same task.
In principle, a measure of the dust should have several advantages over H$\alpha$ and CO. First, the structures can be identified and measured at the same physical scales as the clusters themselves since they are both from the the HST images. Second, we can assess the dust using the same HST images used for the photometry rather than requiring additional 
H$\alpha$ or CO observations. In particular for our project,  
the dust maps are available for all 38 PHANGS-HST galaxies rather than just the 19 galaxies with H$\alpha$ observed as part of PHANGS-MUSE.

Starting with NGC~1433 in the upper left panel of Figure \ref{fig:ha_co_4plot}, 
we find the H$\alpha$ vs. dust  diagram looks relatively similar to the H$\alpha$ vs. CO shown in Figure \ref{fig:co_halpha_4plot}.
There are no young interlopers according to the H$\alpha$ flux (i.e., no blue points above the green line), but five young interlopers according to our dust measurements (the five blue points to the right of the yellow line).
A visual examination again confirms that none of these five are actually young clusters. We note that
%This is similar to Figure \ref{fig:n1433_co_halpha}, where the CO implies there are four interlopers. 
%A manual examination of these objects shows that none of the five are clear interlopers. 
cluster \# 1687 is the same false interloper seen in Figure \ref{fig:7plot-n1433} using CO, but in this case is triggered by a moderate amount of dust  at the edge of the inner star formation region in NGC~1433, rather than by the presence of H$\alpha$ from the MUSE measurement. 
%Similarly, 

In NGC~628 and NGC~3351, the dust measurements are again found to be less reliable than the H$\alpha$ measurements, but similar to CO at identifying young interlopers in the OGC-box (i.e. there are roughly a half dozen more candidates according to the yellow line which are below the H$\alpha$ criteria shown by the green line). 
NGC~1365 shows a large number of young interlopers using both H$\alpha$ and dust as a criteria. 

We also note that in both 
 NGC~3351 and NGC~1365 there is a tendency for the objects with the highest values of H$\alpha$ flux to have low dust values  (i.e., they turn back toward the upper left in Figure \ref{fig:ha_co_4plot}). We believe this is due to a saturation problem since these objects are usually in the densest dust lanes.
 % with clusters with the strongest H$\alpha$ values 
% turning back and having low estimates of dust.
%to the upper left in the figures (i.e. with low values of dust content). 
% These are in the region with the strongest, most saturated dust morphology.  
We are currently investigating potential ways to improve this situation for our dust estimates, but for now we conclude that H$\alpha$ provides the best way to identify interlopers.

\medskip 

{\em Overall, we find that H$\alpha$ emission from MUSE, even at $\approx$factor of 10 lower resolution than the HST images, is better able to identify reddened, young clusters which are interlopers in the OGC-box or the VI-limit region, than CO intensity or relative dust strength}.  This is probably because H$\alpha$ emission is directly tied to the recently formed clusters, whereas molecular gas and dust, which will form the next generation of clusters, has a weaker spatial association with the current generation of young clusters (e.g., \citealp{schruba2010}, \citealp{kruijssen14}, \citealp{kreckel18}, \citealp{kruijssen19}).

Although the focus in this section has been on young interlopers, we note that with a change in the inequality sign (see Sections \ref{sec:approach} and \ref{sec:solutions}), H$\alpha$ (and CO and dust to a lesser extent) can also be used to identify
old interlopers in the high-reddening region of the color-color diagram for NGC 1365.
% as described in Section \ref{sec:solutions}.

\begin{figure*}
\begin{center}
\includegraphics[width =6.6in , angle= 0]{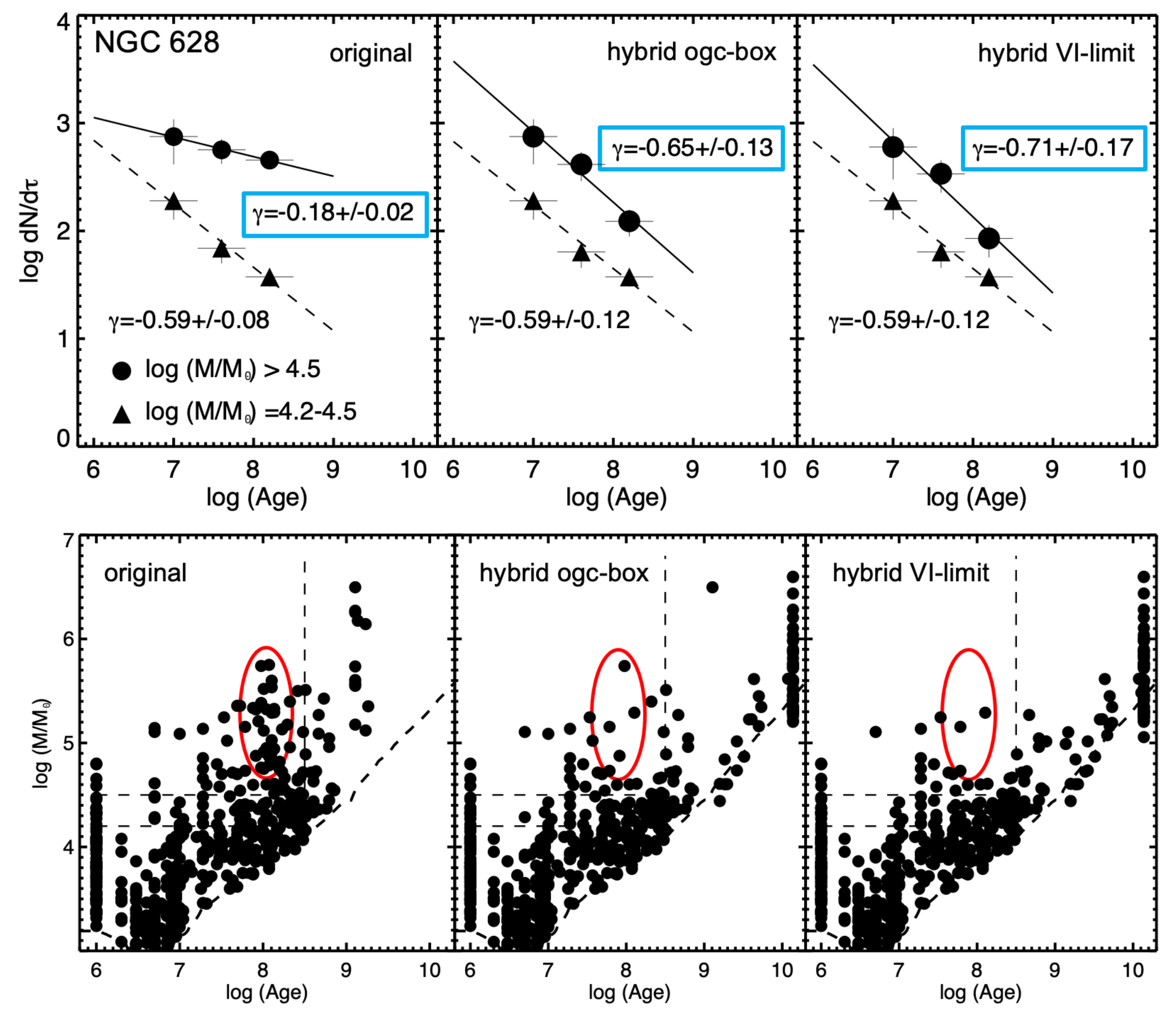}
\end{center}
\caption{PHANGS-HST age distributions in the top panels and log Age vs.log Mass diagrams in the bottom panels for NGC 628. The three panels are, from left to right,  the original, the hybrid OGC-box and the hybrid VI-limit solutions. Two different mass ranges are used as defined by the dashed lines in the bottom plot. The red oval shows the region where most of the bad ages have been corrected. See Figure \ref{fig:7plot-n628} for a color-coded version of the log Age vs.log Mass diagram. Only three age bins are used in the top panel  to facilitate comparison with \citet{adamo17}. }
\label{fig:n628_mass_age}
\end{figure*}

\section{Impact of Incorrect Ages and Masses on Measurements of Cluster Distributions}
\label{sec:age_and_mass} 

\subsection{Age Distribution in NGC~628}
\label{sec:age_and_mass_phangs} 

In the previous sections we show that old globular clusters can often be incorrectly assigned young ages, with moderate to high reddening values, and that reddened young clusters can be mistaken for old globular clusters when standard SED-fitting techniques are applied to cluster populations.  The resulting incorrect ages and masses can potentially impact cluster demographics, especially cluster age distributions.  As an example, in this section we assess the impact on the cluster population in NGC~628, which was included in both the PHANGS-HST and LEGUS surveys.

% In the rest of this section, 
We consider three sets of results: (i) the original age and mass estimates from the PHANGS-HST pipeline, which assume solar metallicity;
(ii) the revised age and mass estimates where young interlopers were identified using the OGC-box; and
(iii) the revised age and mass estimates where young interlopers were identified using the VI-limit approach.

Figure \ref{fig:n628_mass_age}  shows the age distributions in NGC~628 (top panels) for the three different sets of results, and the corresponding age-mass diagrams (bottom panels). The dashed line along the lower edge of the age-mass diagrams shows how the luminosity limit of the sample translates into the age-mass plane.  The dashed vertical and horizontal lines show that the regime used for the age distributions stays above this luminosity limit, so incompleteness should not play a role in the plotted distributions.  The age bins used for each distribution are identical to those presented by \citet{adamo17} for NGC~628 based on the LEGUS survey (more details below).

%Based on looking at the log Age vs.log Mass in Figures *** through ***, 
It is clear from the bottom panels that the three different age-dating methods discussed in this paper impact the overall demographics.  In particular, correcting the ages of clusters found in the OGC-Box, or via the VI-limit,
%, with the removal of many of the blue points (i.e., from within the OGC box) from around 
correctly moves a number of massive clusters originally found to have 
% log~$(\tau/\mbox{yr}) 
log Age $\approx 8$ (i.e., the objects in the red oval) to older ages which are far more appropriate for old globular clusters.  The resulting change in the age distribution is particularly noticeable for clusters more massive than log~$(M/M_{\odot})>4.5$. 

When fit to a power law $dN/d\tau \propto \tau^{\gamma}$, where $\tau$ = Age and $\gamma$ is the power law index,  the original distribution (with incorrect ages for old globular clusters) has a best value of $\gamma=-0.18\pm0.02$ (top-left panel of Figure~\ref{fig:n628_mass_age}).
% essentially identical to the value of $\gamma=-0.19\pm0.07$ found by \citet{adamo17} using the LEGUS data.
However, after either of the two hybrid methods used to identify and correct the ages of red clusters are applied, this distribution is significantly steeper, with $\gamma \approx -0.7$ (top-middle and right panels).  This shows that the determination of the power-law index of the age distribution can be quite sensitive to age-dating results.

Interestingly, the age distribution for lower-mass log~$(M/M_{\odot})=4.2-4.5$ clusters is essentially unaffected by the age corrections (with best fits of $-0.59$ in all 3 cases), at least in NGC~628, because old, globular clusters tend to populate the high mass end around 
% log~$(\tau/\mbox{yr})
log Age $\approx8$ (see Figure \ref{fig:7plot-n628} with color coding).
% The various results are compiled in Table \ref{tab:table2}.

In Figure \ref{fig:n628_dndt2}, we present the age distribution for our best cluster catalog (VI-limit), with double the number of 
% data points 
bins shown in Figure \ref{fig:n628_mass_age}.  Here, we are able to plot the age distribution out to log~Age = 9.2 (above the completeness limit), as opposed to only out to log~Age = 8.4 as used in \cite{adamo17} and Figure \ref{fig:n628_mass_age}, due to the improved treatment of older ages. 
In addition, we include bins at lower ages than \cite{adamo17}; noting that they follow the trend from the older points quite well.
These longer baseline fits also show steep  power law fits for the hybrid age-dating solutions, with $\gamma=-0.78\pm0.07$ for the high mass fit and $\gamma=-0.67\pm0.10$ for the low mass fit.
If we remove the youngest age bin from the fit, due to  concerns about potential inclusion of unbound associations \citep{bastian12b}, the fits are only weakly affected, with $\gamma=-0.74\pm0.07$ for the high mass fit and $\gamma=-0.50\pm0.05$ for the low mass fit.
%(see Table \ref{tab:table2}). 

\subsection{Comparison with LEGUS Results for NGC~628}
\label{sec:age_and_mass_legus} 

%**** CONDENSE BELOW TO A SINGLE PARA ***

\begin{figure}
\begin{center}
\includegraphics[width =3.3in , angle= 0]{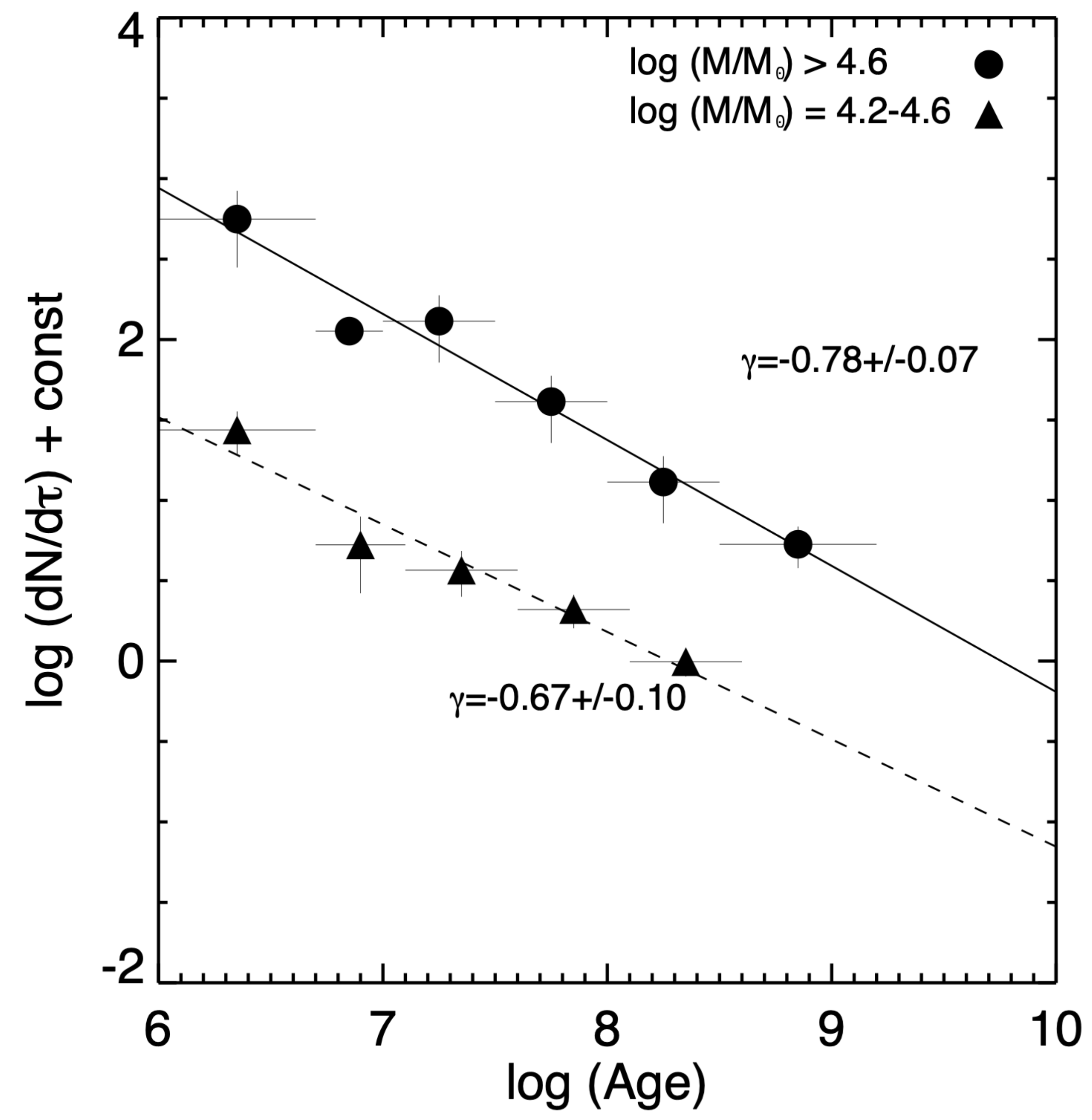}
\end{center}
\caption{Age distributions determined from the (best) VI-limit cluster catalog.  These show  a significantly larger number of data points than the previous figure
% can be used once the corrections to the older age clusters have been made 
(see Section \ref{sec:age_and_mass_phangs} for further discussion).}
\label{fig:n628_dndt2}
\end{figure}

The cluster population in NGC~628 was studied previously by the LEGUS project; see \cite{grasha15} and \cite{adamo17}.  That collaboration used the same HST data (pointings and set of five filters) as used in this work, but use the Yggdrasil SED models \citep{zackrisson11}.
Several different combinations of SEDs, extinction laws and aperture corrections are provided in the  LEGUS study. We have chosen to use the Padova isochrones, a Milky Way extinction law, and mean aperture corrections. The results do not vary much for the other combinations. 
The cluster properties discussed here can be directly compared with our own result from Figures \ref{fig:n628_mass_age} and \ref{fig:n628_dndt2},  and Section \ref{sec:age_and_mass_phangs}. 

We find that 134 out of 864 (16\%) class 1$+$2 human-classified LEGUS clusters are found in the 
%LEGUS human classified objects for class 1 and 2 clusters, there are 134 clusters (blue points) in the 
OGC-box.
%out of a total of 864 (i.e., 16 \%). 
This is similar to the 13\% we find in the original PHANGS-HST pipeline results, as listed in Table \ref{tab:table1}. 
The log Age vs. $E(B-V)$ diagram for LEGUS (not shown)  is very similar to the upper left panel of our Figure \ref{fig:7plot-n628}.
% The top panel shows that %like our own result for NGC~628 in Figure \ref{fig:7plot-n628}, 
Nearly all of the clusters in the OGC-Box for the LEGUS results are above the diagonal, just like the results we presented earlier for the original PHANGS-HST pipeline age dating. In addition, there is 
only one object with log Age $>$ 9.5, similar to the zero objects with log Age $>$ 9.5 for the original PHANG-HST ages. This indicates that the LEGUS age-mass results have the same problems as the PHANGS-HST pipeline catalogs, which is reasonable since they adopted  
%% maximum $E(B-V)$ during their SED fitting and 
similar fitting techniques and maximum reddening. This also suggests that the effects on their science results will be similar to those discussed in Section \ref{sec:age_and_mass_phangs}.

\cite{adamo17} find a shallow power-law index for their (Class $1+2$) cluster age distribution, $-0.19 \pm 0.07$,  based on the same three age intervals shown in the bottom panels of Figure~\ref{fig:n628_mass_age}.
%vs. $-0.49 \pm 0.23$ for our high mass and $-0.68 \pm 0.04$ for our low mass. 
We found essentially identical results at the high mass end (notwithstanding the systematic $\sim$0.4~dex difference in mass estimates) when we use the original PHANGS-HST catalog (top-left panel of Figure~\ref{fig:n628_mass_age}, which made similar assumptions during the age-dating procedure.

%Notwithstanding the apparent difference in mass estimates between LEGUS and PHANGS-HST, 
%The main conclusion is 
Hence, we find that the original PHANGS-HST log Age vs. $E(B-V)$ diagram, as well as the original PHANGS-HST log Age vs. log Mass and log Age vs. dN/dT diagrams,  all look essentially identical to the  LEGUS versions, showing  that both studies are affected by the age/metallicity/degeneracy problem. Fixing this problem using the hybrid solutions introduced in the current paper results in steeper age distributions due to correctly moving $\approx30$ old globular clusters from their original ages of 
% $log~$(\tau/\mbox{yr}) 
log~Age $\approx 8$ to ages log~Age $> 9.5$.

The corrected age distributions for clusters  in NGC~628 have important implications for the disruption of the clusters.   We find that these distributions decline more-or-less continuously starting at very young ages, and have a similar shape for clusters at different masses.  The simplest interpretation of this result is that the initial masses of the clusters do not strongly impact their dissolution time, often referred to as mass-independent disruption \citep{fall12,bastian12b}.  However, this is not a unique interpretation. Simulations using the MOSAICS framework \citep{kruijssen11} have shown that the disruption rate of clusters changes over time and with location, as clusters migrate from more to less disruptive environments after they form.  Including these variations in the disruption rate often also results in a steep age-distribution, with only a weak dependence on the cluster mass, even when mass-dependent cluster disruption models are adopted \citep{kruijssen11, Miholics17}.   In any case, it is critical that observational works accurately establish the shape of the cluster age distribution in different galaxies, and in particular that they correct mistakes in the age-dating of old globular clusters.

\subsection{Impact on Other Works}

Our results make it clear that contamination by globular clusters may have had a significant impact on previous studies of cluster populations in star-forming galaxies. 
We have examined several previous papers on cluster demographics and found that many others were affected, at least at some level, by the same age/reddening/metallicity degeneracy challenges discussed in this work. These include \cite{whitmore99}, \cite{hunter03}, \cite{larsen04}, \citealp{fall05}, \cite{gieles05}, \cite{whitmore10}, \cite{bastian12b}, \cite{adamo17}, 
% \cite{messa21}, 
and  \cite{moeller22}, to list  a few. 
This is not meant to be a comprehensive list; we expect that %essentially all 
most works that rely on the results of age-dating stellar clusters that do not include H$\alpha$ or some other narrow-band measurement directly in the fits  may be affected by these issues at some level.  A systematic technique to identify ancient globular clusters using multi-band SED fits will be important going forward.
%problem at some level. 
%As summarized in Section \ref{sec:conclusions}, 
We recommend that in the future, authors consider including the following graphics in their manuscripts to make it clear to what degree their results might be affected by the age/reddening/metallicity degeneracies: (1) log Age vs. $E(B-V)$ diagrams, (2) color-color diagrams, (3) log Age vs. log Mass diagrams, (4) and color images of a few relevant objects.

Many early papers, like \cite{whitmore99} observations of the Antennae galaxies, relied on the WFPC2 camera for U-band photometry. The poor quantum efficiency of this instrument at shorter wavelengths, roughly a factor of ten less than for WFC3 at F336W, limited the scope of the problem because most old globular clusters have sufficiently faint U-band fluxes that they fell out of the samples.  However, with the significant improvement to the blue imaging capabilities with the ACS/HRC and WFC3 cameras, many globular clusters now have measurements at shorter wavelengths which can lead to the problems highlighted in the current work.  

%In Section \ref{sec:age_and_mass} 
Earlier in this Section, we described how incorrect age-dating can affect the steepness of the age distribution for star clusters.
The preferential impact this problem appears to  have  on the  high mass end (see Figure \ref{fig:7plot-n628} and the discussion in Section \ref{sec:age_and_mass_phangs}) could potentially 
also affect the conclusion of whether 
cluster destruction is mass-dependent or mass-independent, as discussed in a variety of papers (e.g.,  \citealp{bastian12b}, \citealp{fall12}, \citealp{krumholz15}).
Other potential results that may be affected are whether the upper end of the cluster mass function is best described by a power law or by a Schechter function (e.g., \citealp{fall01},  \citealp{gieles09}, \citealp{adamo15},  \citealp{Messa18}, \citealp{mok19}), and estimates of the fraction of stars that are born in clusters, commonly known as $\Gamma$ (\citealp{bastian08},  \citealp{kruijssen12}, \citealp{adamo15}, \citealp{chandar17}), among other important results.  
%For NGC~628, we estimate that $\Gamma \sim31$\% from the original PHANGS-HST pipeline age-dating (measured from $1-10$~Myr clusters). Using the hybrid VI-limit ages instead lowers $\Gamma$ to  $\sim24$\%.  
These and other issues related to the effect of using the improved hybrid ages will be quantified in upcoming papers (Mok et al, and Chandar et al., in preparation).

\section{Future Work}
\label{sec:future_work}

The current study is a pilot program that will help improve the PHANGS-HST pipeline results for cluster ages, masses, and reddening. 
The solutions developed as part of the current paper are included as online MNRAS data products associated with this article.  
The original version 1 catalogs \citep{turner21}, as well as those developed from the future PHANGS-HST pipeline,  will all be included in the PHANGS-HST archives at \url{https://archive.stsci.edu/hlsp/phangs-cat}.

The techniques developed in the current paper might be described as a "post-facto" approach, that is, running multiple  CIGALE models with a variety of metallicities and $E(B-V)$ limits, and then using other information (i.e., colors, CO intensities, H$\alpha$ flux, dust) to decide which solution is most appropriate for a given star cluster "after the fact". We are currently developing a more  "ab-initio" approach, where priors, partially developed from the current study, will be used by the CIGALE software early in the process to select the correct peak (age/reddening/metallicity combination) in a multi-modal probability distribution
%of some clusters in earlier stages of the process 
(see Figure \ref{fig:bimodal}).

One advantage of the ab-initio approach is that it will seamlessly make corrections for all the clusters, not just the ones with the most obvious problems (i.e., old globular clusters, heavily reddened clusters), as discussed in the current paper. Similarly, in the current approach there are discontinuities at the edges of regions identified in color-color space that can be eliminated. 

One of the primary problems for the current project was the poor resolution of the CO and MUSE H$\alpha$ observations, as discussed in Section \ref{sec:solutions}. H$\alpha$ observations using the Hubble Space Telescope for the missing galaxies will significantly improve this aspect (i.e., proposal 17126 - PI = Chandar). 
Similarly, improvements in the newly developed dust maps (e.g., Thilker et al. 2022) might also make this parameter more useful for measuring ages.  In principle, it may be possible to use a wide variety of information (e.g., H$\alpha$, CO, dust, $E(B-V)$ Balmer decrement, etc.) simultaneously, rather than just H$\alpha$.
In addition, we plan to make extensive tests of the effects of a variety of different parameters including metallicity, reddening laws, and the presence of binaries and horizontal branch stars, to name just a few.

Finally, results which include near-infrared photometry from JWST observations \citep{lee22b}
% PI = J. Lee, proposal-ID = 2107) 
will be presented in future catalogs. This should be especially valuable for the population of deeply embedded clusters in dusty galaxies like NGC~1365, and will also help with age-dating since PAH emission is primarily associated with young regions (\citealp{papovich07}, \citealp{popescu11}).

\section{Summary and Conclusions}
\label{sec:conclusions} 

New techniques have been developed to improve SED age estimates of star clusters for four galaxies from the PHANGS-HST survey. The project is designed as a pathfinder for the future development of an improved PHANGS-HST pipeline, and will eventually incorporate new measurements from JWST observations. The primary focus is to reduce or eliminate effects of the age/reddening/metallicity degeneracy for the two most pressing problems: 1) old globular clusters with age estimates that are too young (and reddening that is too high), and 2) highly reddened young clusters with age estimates that are too old (and reddening that is too low).

 A problem with most current age-dating studies is the use of a single metallicity, generally solar.  This results in a mismatch between the observations and SED models that invalidates the fitting process at the first step for many objects. In particular, old globular clusters have low metallicity that puts them above the solar SED models in a $U-B$ vs. $V-I$ diagram.
 The primary strategy used in the current paper is to identify likely old globular clusters from their position in a color-color diagram, and use 1/50th metallicity models with $E(B-V)$ constrained to be less than 0.1 mag.
 %to determine the age/reddening/mass. 
 Interlopers (generally young objects reddened by dust rather than age) are identified using H$\alpha$ flux information and are given the original solar metallicty solution.  CO intensities and dust maps were also examined for this purpose, but found to be less effective.

The CIGALE software \citep{Boquien19} is used with Bruzual-Charlot models as the primary software tool.  The observational inputs are five color (F275W, F336W, F435W, F555W, F814W) HST photometry from Version 1.1 of the PHANGS-HST pipeline \citep{turner21}, and  H$\alpha$ flux values from MUSE \citep{emsellem22}. Secondary information is provided in the form of CO observations from ALMA (as part of the PHANGS consortium \citealp{leroy21}), and dust models from \cite{thilker23}.

The primary results are:

1. Age-dating success fractions 
%% (i.e., no differences between manually estimated ages and predicted ages larger than a factor of 10)  
are improved for the program galaxies by approximately 9~\% (NGC 628), 9~\% (NGC 3351), 13~\% (NGC 1365),  and  18~\% (NGC 1433).
Success is defined as limiting differences between manually estimated and predicted ages to less than a factor of ten. While these may seem to be relatively small improvements, for certain objects (e.g., old globular clusters) the difference can be 100 \%. For example, the increase in the number of old globular clusters (i.e., log Age > 9.5) goes from 0  to 65 for NGC 628. In addition, the majority of the incorrect ages for old globular clusters are systematically assigned values in the range 5 to 500 Myr. This can pollute the study of young and intermediate age cluster populations, impacting a variety of key science results.
%and hence affect a variety of science results. 

2. Two approaches are used for the three relatively dust-free galaxies (NGC 1433, NGC 3351, NGC 628). The first uses positions in the $U-B$ vs. $V-I$ diagram to identify potential Old Globular Clusters (i.e., the OGC-box approach). The second uses the VI-limit approach, so that U-band measurements are not required. The VI-limit approach appears to be better in all three  galaxies, with  overall success fractions of approximately 97~\%.
 This method is expected to work well for all but a few of the dustiest PHANGS-HST galaxies.

3. A somewhat different approach is used for the very dusty, high SFR galaxy NGC~1365.
The primary differences are the use of a smaller OGC-box, and the inclusion 
of an additional high-reddening region of the color-color diagrams
(i.e., where young clusters are often given incorrect estimates of old ages and low reddening - opposite to the problem with old globular clusters in the OGC-box). The overall success fraction for the new hybrid age estimates for NGC 1365 is approximately  92\%.

4.  H$\alpha$ emission currently provides the best method of identifying and correcting "interlopers" from the default procedures outlined above (i.e. young interlopers in the OGC-box or VI-limit approach; old interlopers in the high-red region for NGC 1365).
% %, and insuring they get the correct age. CO intensities and dust strength are roughly equally effective, but not as good at  identifying interlopers correctly as H$\alpha$. 
This may be because H$\alpha$ emission is directly tied to the recently formed clusters, whereas molecular gas and dust has a weaker spatial association with the current generation of young clusters. 

5. An interesting aspect of NGC~1365 is a large number of $\approx300$~Myr clusters, and a reddening vector `plume' of points from this population clearly visible in the $U-B$ vs.$V-I$ diagram. We believe this is a natural result of extensive star formation along the bar which eventually separates itself from the gas and dust which tends to spiral into the central region due to dissipation.

6. The errors in age estimates from the age/reddening/metallicity degeneracy result in systematic rather than random errors. As such, they can affect a wide variety of science results. To demonstrate, we examine the power law index of the age distribution in NGC 628 and find it steepens by approximately 0.5 for high mass clusters. The improved ages also allow us to push to larger values of log Age, and hence increase the number of data points used in fits for various correlations.

\medskip

The problems described in this paper are potentially endemic to a large number of studies from the past several decades. They are especially prevalent after WFC3 became available  in 2010 with its excellent UV sensitivity and wide field of view. Before this,  
most old globular clusters fell out of age-dating star cluster samples in external galaxies  due to the faint UV flux from their old populations. The  effects of these problems may be quite widespread. 
Examples include the  steepness of the age functions (e.g., see Section \ref{sec:age_and_mass}), the apparent cutoff at the high end of the mass function, the fraction of stars that form in clusters, and the specific frequency of old globular clusters in galaxies. 

Potential evidence of the effects of the age/reddening/metallicity degeneracy in a sample of star clusters include: 1) the lack of old globular clusters in a sample (i.e., few or none beyond log Age = 9.5), and 2) a large number of intermediate-age  clusters with high values of reddening (i.e., above a `diagonal' running from log Age, $E(B-V)$ = (6.0, 1.0) to (9.0, 0.1)) .
%% 3) a large number of objects in parts of the color-color diagrams usually associated with old globular clusters but which are given intermediate or young ages. 
We encourage authors to include diagrams such as color-color, log Age vs. $E(B-V)$, and log Age vs. log Mass plots to help
identify the effects of this problem in different studies. 

% The original cluster catalogs, as well as those  with revised ages created as part of this effort are available through the PHANGS-HST archives at \url{https://archive.stsci.edu/hlsp/phangs-hst}. Additional catalogs will also be developed in the future for the remaining 15 PHANGS-HST galaxies that have MUSE IFU spectra available.

The original version 1 cluster catalogs developed by \cite{turner21} are available through the PHANGS-HST archives at \url{https://archive.stsci.edu/hlsp/phangs-hst}. The catalogs developed as part of the current paper are included as online MNRAS data products associated with this article.    
%Additional catalogs will also be developed in the future for the remaining 15 PHANGS-HST galaxies that have MUSE IFU spectra available, and are available from the author on request. 
The PHANGS-HST pipeline will use the lessons learned in this pilot study to develop an "ab-initio" approach within CIGALE, selecting peaks from the multi-modal probability distributions, as briefly described in Section \ref{sec:future_work}. These new catalogs will  be included in the PHANGS-HST archives at \url{https://archive.stsci.edu/hlsp/phangs-cat}.

\bigskip

\section*{Acknowledgements}

We thank the referee for several useful and constructive comments that lead to improvements in the paper.
Based on observations made with the NASA/ESA Hubble Space Telescope, obtained from the data archive at the Space Telescope Science Institute. STScI is operated by the Association of Universities for Research in Astronomy, Inc. under NASA contract NAS 5-26555.  Support for Program number 15654 was provided through a grant from the STScI under NASA contract NAS5-26555. FB and AB would like to acknowledge funding from the European Research Council (ERC) under the European Union’s Horizon 2020 research and innovation programme (grant agreement No.726384/Empire).
MB gratefully acknowledges support by the ANID BASAL project FB210003 and from the FONDECYT regular grant 1211000. 
EW acknowledges support from the DFG via SFB 881 ‘The Milky Way System’ (project-ID 138713538; subproject P01).
FB acknowledges funding from the European Research Council (ERC) under the European Union’s Horizon 2020 research and innovation programme (grant agreement No.726384/Empire).
HAP acknowledges support by the National Science and Technology Council of Taiwan under grant 110-2112-M-032-020-MY3.
JMDK acknowledges funding from the European Research Council (ERC) under the European Union's Horizon 2020 research and innovation programme via the ERC Starting Grant MUSTANG (grant agreement number 714907). COOL Research DAO is a Decentralised Autonomous Organisation supporting research in astrophysics aimed at uncovering our cosmic origins.
% JMDK gratefully acknowledges funding from the Deutsche Forschungsgemeinschaft (DFG, German Research Foundation) through an Emmy Noether Research Group (grant number KR4801/1-1), as well as from the European Research Council (ERC) under the European Union's Horizon 2020 research and innovation programme via the ERC Starting Grant MUSTANG (grant agreement number 714907).
KG is supported by the Australian Research Council through the Discovery Early Career Researcher Award (DECRA) Fellowship DE220100766 funded by the Australian Government.
KK and FS gratefully acknowledge funding from the German Research Foundation (DFG) in the form of an Emmy Noether Research Group (grant No. KR4598/2-1, PI Kreckel).
PSB acknowledges financial support from the Spanish Ministry of Science, Innovation and Universities under grant number PID2019-107427GB-C31.
 RSK and MCS are thankful for support from the Deutsche Forschungsgemeinschaft (DFG) via the Collaborative Research Center (SFB 881, Project-ID 138713538) ``The Milky Way System'' (sub-projects A1, B1, B2 and B8) and from the Heidelberg Cluster of Excellence (EXC 2181 - 390900948) ``STRUCTURES: A unifying approach to emergent phenomena in the physical world, mathematics, and complex data'', funded by the German Excellence Strategy. RSK and MSC also acknowledge funding from the European Research Council in the ERC Synergy Grant ``ECOGAL -- Understanding our Galactic ecosystem: From the disk of the Milky Way to the formation sites of stars and planets'' (project ID 855130). TGW acknowledges funding from the European Research Council (ERC) under the European Union’s Horizon 2020 research and innovation programme (grant agreement No. 694343).

\section{Data Availability}

The imaging observations underlying this article can be retrieved from the Mikulski Archive for Space Telescopes at \url{https://archive.stsci.edu/hst/search_retrieve.html} under proposal GO-15654. High level science products, including science ready mosaicked imaging, associated with HST GO-15654 are provided at \url{https://archive.stsci.edu/hlsp/phangs-hst} with digital object identifier \doi{10.17909/t9-r08f-dq31}

\clearpage

%%%%%%%%%%%%%%%%%%%%%%%%%%%%%%%%%%%%%%%%%%%%%%%%%%

 \begin{table}
 % \caption{Success Statistics for the Four Program Galaxies using H$\alpha$ to Identify Interlopers}
 \caption{Statistics for the Four Program Galaxies}

 \label{tab:table1}
 \begin{tabular}{lccccc}
  \hline
 Galaxy (star formation rate) & N 1433 (SFR = 0.56)      & N 3351 (SFR = 0.87) & N 628 (SFR = 0.93) & N 1365 (SFR = 16.9) \\
  \hline
  \\
  number of class 1 and 2 clusters (human classification) & 191 & 317 & 489 & 635 \\
  
  number in OGC-Box (mainly old globular clusters)  & 49/191 = 26 \% & 56/317 = 18 \%  & 63/489 = 13 \% & 88/635 = 14 \%  \\
  
  number in VI limit region & 58/191 = 40 \% & 75/317 = 24 \%  & 96/489 = 20 \% &  \\
  
  number in high red region (mainly young, dusty)  &  &  &  & 43/635 = 7 \% \\
  
  \\
  
  standard solution, above diagonal (i.e., ages suspect) & 40/191 = 21 \% & 41/317 = 13 \% & 70/489 = 14 \% & 128/635 = 20 \% \\
  
  hybrid OGC-box solution,  above diagonal (i.e., ages suspect)  & 3/191 = 2 \% & 12/317 = 4 \% & 23/489 = 5 \% &  \\
  
  hybrid VI-limit solution,  above diagonal (i.e., ages suspect)   & 0/191  = 0 \% & 7/317 = 2 \%  &  12/489 = 2 \% &  \\
  
  NGC1365 solution,  above diagonal (i.e., ages suspect)   & &  & & 120/635 = 19 \% \\
  
  \\
  standard solution, old globular clusters (log Age $>$ 9.5)     & 1/191  = 1 \% & 7/317 = 2 \%  & 0/489 = 0 \% & 22/635 = 3 \% \\
  
  hybrid OGC-box solution, old globular clusters 
  log Age $>$ 9.5)     
  & 43/191  = 23 \% & 44/317 = 14 \%  &  53/489 = 11 \% &  \\
  
  hybrid VI-limit solution, old globular clusters (log Age $>$ 9.5)     & 47/191  = 25 \% & 50/317 = 16 \%  &  65/489 = 13 \% &  \\
  
  hybrid (NGC1365), old globular clusters (log Age $>$ 9.5)     &   &   & & 55/635 = 9 \% \\

  \\
  \\
  
  Reason remaining above diagonal using hybrid approach:
  
  \\
  \\
  
  case \# 1 - Unreliable U-band - GOOD/BAD$^a$ age & OGC: 3/0, VI: 0/0 & OGC: 0/4, VI: 0/0 & OGC$^b$: 3/8, VI$^b$: 1/1 & 4/1 \\
  
  \\
  
  case \# 2 -  OGC-box interloper:  - GOOD/BAD age & OGC: 0/0, VI: 0/0 & OGC: 3/3, VI: 3/2 & OGC: 1/0, VI: 1/0 & 14/4 \\
  
  \\
  
  case \# 3 - VI-limit interloper:   - GOOD/BAD age & OGC: 0/0, VI: 0/0 & OGC: 0/0, VI: 0/0 & OGC: 1/0, VI: 1/0 & 3/2  \\
  
  \\
  
  case \# 4 - High red (N1365):   - GOOD/BAD age  &  & & & 32/7\\

  \\
  
  case \# 5 - $~$ 300 Myr plume (N1365): - GOOD/BAD age  &  & & & 34/4  \\

  \\
  
  case \# 6 - Resolution problem- GOOD/BAD age & OGC: 0/0, VI: 0/0 & OGC: 0/1, VI: 0/1 & OGC: 0/0, VI: 0/0 & 3/2   \\
  
  \\
  
 %  case \# 7 - Artifact - GOOD/BAD age & OGC: 0/0, VI: 0/0 & OGC: 0/1, VI: 0/0 & OGC: 0/0, VI: 0/1 & 0/1  \\
  
  case \# 7$^c$  - Artifact - GOOD/BAD age & OGC: 0/0, VI: 0/0 & OGC: 1/1, VI: 1/0 & OGC: 7/2, VI: 6/3 & 10/5  \\
  
  \\
  
  % case \# 8 - Misc - GOOD/BAD age & OGC: 0/0, VI: 0/0 & OGC: 1,0, VI: 1,0 & OGC: 6/2$^b$, VI: 6/2$^b$ & 7/3 \\
 
 % \\
  
  % case \# 9 stochastic - GOOD/BAD age & OGC: 0/0, VI: 0/0 & OGC: 0/0, VI: 0/0 & OGC: 1/0  VI: 0/1 & 0/0\\
  
 % \\

  Summary - OGC-box (remaining BAD ages above diag.) & 3/191 = 2 \%  & 9/317 = 3 \% &  10/489 = 2 \%
  \\
  
  Summary - VI-limit (remaining BAD ages above diag.) & 0/191 = 0 \% & 3/317 = 1 \%  &  5/489  = 1 \% &  \\
  
  Summary - N1365  (remaining BAD ages above diag.) &  & & & 24/635 = 4 \% \\
  
  \\
  
  Success \%  (assume 96 \% and 90 \% [n1365] below diag.)  & 97 \%  & 97 \% & 97 \% & 92 \%  
  \\

 \hline
 \end{tabular}
 $^a$ - Evaluation of whether good or bad age estimate (i.e., within a factor of ten). Based on a visual review of the B-V-I, H$\alpha$, and CO images.\\
 $^b$ - Unreliable U-band because off the edge of the image. \\
 $^c$ - Artifacts include cluster misclassifications (e.g., slightly saturated stars, close pairs of stars, background galaxies, ...), objects with colors that place them just outside of the color criteria so no correction is made, but should have been made (e.g., $V-I$  = 0.94), objects where the solution reverts to the original solar metallicity solution and that turns out to be wrong (e.g., an object with strong H$\alpha$ given an age of  100 Myr).
 
\end{table}

\clearpage


\bibliographystyle{mnras}   
% \bibliography{phangshst}  
% \bibliography{all}  
\bibliography{all_nov_24_2021}  
% \bibliography{phangsjwst}  

%%%%%%%%%%%%%%%%%%%%%%%%%%%%%%%%%%%%%%%%%%%%%%%%%%

%%%%%%%%%%%%%%%%% APPENDICES %%%%%%%%%%%%%%%%%%%%%

%\appendix
%\section{TBD}

% Don't change these lines
\bsp	% typesetting comment
\label{lastpage}
\end{document}